\documentclass[12pt, oneside]{article}
\usepackage[utf8]{inputenc}
\usepackage{appendix}
\usepackage{geometry} 
\usepackage{amsmath} 					\usepackage{fancyhdr}			
\usepackage{graphicx}	
\usepackage{amssymb}
\usepackage{cite}
\usepackage{color,xspace}
\usepackage{extarrows}
\usepackage{mathtools}
\usepackage{comment}
\usepackage{braket}
\usepackage[skip=2pt,font=scriptsize]{caption}

\numberwithin{equation}{section}

\label{key}
\usepackage[margin=1cm]{caption}

\DeclareSymbolFont{extraup}{U}{zavm}{m}{n}
\DeclareMathSymbol{\vardiamond}{\mathalpha}{extraup}{87}
\usepackage{float}
\restylefloat{table}
\allowdisplaybreaks

\def\twomat[#1,#2][#3,#4]{\left( \begin{array}{cc} #1 & #2 \\ #3 & #4 \end{array} \right)}
\def\thv[#1,#2,#3]{\left( \begin{array}{c} #1 \\ #2 \\ #3 \end{array} \right)}
\def\twv[#1,#2]{\left( \begin{array}{c} #1 \\ #2 \end{array} \right)}

\title{The simplest }

\date{}

\begin{document}
	
	\begin{flushright}
	\end{flushright}
	\begin{center}
		
		\vspace{1cm}
		{\LARGE{\bf Newton versus Coulomb for Kaluza-Klein modes}}
		
		\vspace{1cm}

		\large{\bf Karim Benakli$^a$ \let\thefootnote\relax\footnote{$^a$kbenakli@lpthe.jussieu.fr},
			Carlo Branchina$^b$ \let\thefootnote\relax\footnote{$^b$cbranchina@lpthe.jussieu.fr}
			and
			Ga\"etan~Lafforgue-Marmet$^c$ \footnote{$^c$glm@lpthe.jussieu.fr}
			\\[5mm]}
		
		{ \sl Sorbonne Universit\'e, CNRS, Laboratoire de Physique Th\'eorique et Hautes Energies, LPTHE, F-75005 Paris, France.}

	\end{center}
	\vspace{0.7cm}
	
	\abstract{ 
		We consider a set of elementary compactifications of $D+1$ to $D$ spacetime dimensions on a circle: first for pure general relativity, then in the presence of a scalar field, first free then with a non minimal coupling to the Ricci scalar, and finally in the presence of gauge bosons. We compute the tree-level amplitudes in order to compare some gravitational and non-gravitational amplitudes. This allows us to recover the known constraints of the $U(1)$, dilatonic and scalar Weak Gravity Conjectures in some cases, and to show the interplay of the different interactions. We study the KK modes pair-production in different dimensions. We also discuss the contribution to some of these amplitudes of the non-minimal coupling in higher dimensions for scalar fields to the Ricci scalar.
	}

	\newpage
	
	\tableofcontents
	
	\setcounter{footnote}{0}
	
	%------------------------------------------------------------------------------------------------------------							
	%\section{Introduction}
	%\label{introduction}
	
	%------------------------------------------------------------------------------------------------------------

	%%%%%%%%%%%%%%%%%%%%%%%%%%%%%%%%%%%%%%%%%%%%%%%%%%%%%%%%%%%%%%%%%%%%%%%%
	%
	%
	%%%%%%%%%%%%%%%%%%%%%%%%%%%%%%%%%%%%%%%%%%%%%%%%%%%%%%%%%%%%%%%%%%%%%%%%
	
	%
	%\title{Kaluza Klein Factory}
	%\author{}
	%\date{November 2020}
	%
	%\begin{document}
	%
	%\maketitle
	
	\section{Introduction }

	Among the Swampland conjectures \cite{Vafa:2005ui}, one of the most popular and best tested is probably the Weak Gravity Conjecture (WGC). Its simplest formulation \cite{ArkaniHamed:2006dz} considers the case of a $D$-dimensional $U(1)$ gauge theory, with a coupling constant $g$, and requires the existence of at least one state of mass $m$ and charge $q$ which satisfies:
	\begin{equation}
		\label{WGC}
		gq\ge\sqrt{\frac{D-3}{D-2}}\kappa_D m,
	\end{equation}
	where $\kappa_D$ is defined as $\kappa^2_D=8\pi G_D=\frac{1}{M_{P,D}^{D-2}}$ with $M_{P,D}$ the reduced Planck mass in $D$ dimensions. This inequality implies, among others, that in the non-relativistic limit, the Newton force is not stronger than the Coulomb force. The particular states for which the equality in \eqref{WGC} is satisfied are said to saturate the WGC.  In this work we will be interested in a particular case of them.

	The present work is dedicated to the study of two different generalizations of the WGC: one that arises when the gauge interaction is complemented by a dilaton interaction\cite{Heidenreich:2015nta,Benakli:2021fvv}, and another \cite{Gonzalo:2019gjp,Benakli:2020pkm,Gonzalo:2020kke} that broadly requires the dominance of scalar interactions with respect to gravity in some scattering processes depending on the specific theory. We are interested in  the modes that propagate in an extra dimension forming a tower of KK excitations  \cite{Kaluza:1921tu,Klein:1926tv,Klein:1926fj,Einstein:1938fk}.  We will explicitly show that these modes undergo gravitational and non-gravitational interactions of equal intensity, which allows us to use them as probes for the conjectured inequalities generalizing the one mentioned above. They will also be useful to investigate the behavior of the scalar WGC under compactification.  
	
	Obviously, the KK excitations considered here saturate the inequalities conjectured only at the classical level, to which our study will be limited, since both terms of these inequalities are in general corrected by quantum effects. However, one has in mind that extending the theory with enough supersymmetries, the KK modes can be BPS states which saturate them even at the quantum level.
	
	The fact that KK modes saturate the inequalities of the various conjectures is a known property, but we will give a derivation of it here in a simple form that we have not found in the existing literature. Our derivation of the various inequalities will be based on amplitude calculations, not for example on the conditions for decay of extremal black holes, and some of the explicit expressions for the amplitudes needed to make the comparisons seem to be either missing or scattered and hard to find, so we hope that presenting them altogether here might be useful.
	
	This work is organized as follows. Section 2 reviews the well-known reduction of KK from $D+1$ to $D$ dimensions of the Hilbert-Einstein action and a massless scalar. It allows us to introduce our notations, presents the Lagrangian expansion needed to extract the Feymann rules for calculating amplitudes, and compute the numerical  factor in the total derivative term, often misquoted in the literature, which will be useful in Section 5. The dilatonic WGC inequality is derived in Section 3, where we also calculate various KK pair production amplitudes.  In Section 4, we consider adding a mass term for the scalar in $D+1$ dimensions and we find our form of the scalar WGC. A non-minimal coupling to gravity is considered in section 5. The interactions due to the presence of higher dimensional gauge fields are discussed in section 6. Our conclusions are presented in section 7. Finally, some technical details about our calculations are gathered in appendices.

	%%%%%%%%%%%%%%%%%%%%%%%%%%%%%%%%%%%%%%%%%%%%%%%%%%%%%%%%%%%%%%%%%%%%%%%%%%%%%%%%%%%%%%%%%%%%

	\section{Expansion to Second Order in the Gravitational Field}
	
	%\subsection*{Notation}
	We work with the signature $(+,-,...,-)$. The $D+1$ dimensional quantities will be denoted with a hat. We use Latin and Greek letters  for the D+1 and D-dimensional coordinates, respectively. We denote by $x$ the $D$ non-compact and by $z \equiv z + 2 \pi L$ the compact coordinates. We recall the steps of the simple dimensional reduction of a free real massless scalar field $\hat\Phi$ coupled to General Relativity:
	\begin{equation}
		\mathcal{S}^{(D+1)}=\mathcal{S}_{EH}^{(D+1)}+\mathcal{S}_{\Phi,0}^{(D+1)},
		\label{free scalar action}
	\end{equation}
	where 
	\begin{equation}
		\label{action}
		\mathcal{S}_{EH}^{(D+1)}=\frac{1}{2\hat{\kappa}^2}\int\mathrm{d}^{D+1}x \sqrt{(-1)^D\hat{g}}\,\hat{R},
	\end{equation}
	and 
	\begin{equation}
		\label{scalar action}
		\mathcal{S}_{\Phi,0}^{(D+1)}=\int\mathrm{d}^{D+1}x\,\,\sqrt{(-1)^{D}\hat{g}}\,\,\frac{1}{2}\hat{g}^{MN}\partial_M \hat\Phi \partial_N \hat\Phi
	\end{equation}
	
	The Ricci scalar $\hat{R}$ is computed from the metric $\hat{g}_{MN}$. In the simplest compactification from  $D+1$ to $D$ dimensions it takes the form 
	\begin{equation}
		\label{metric}
		\hat{g}_{MN}=\begin{pmatrix}
			e^{2\alpha\phi}g_{\mu \nu}-e^{2\beta \phi}A_{\mu}A_{\nu} & e^{2\beta \phi}A_{\mu} \\ 
			e^{2\beta \phi}A_{\nu} & -e^{2\beta \phi}
		\end{pmatrix}
	\end{equation}
	with $\phi$, $A_{\mu}$ and $g_{\mu \nu}$ D-dimensional fields independent of the $z$ coordinate:
	
	\begin{align}
		\label{actionbeforespecifyingabeta}
		\mathcal{S}_{EH}^{(D+1)}=\frac{1}{2\hat{\kappa}^2} \int \mathrm{d}^{D+1}x\, \sqrt{(-1)^{D-1}g} \,\, e^{((D-2)\alpha+\beta)\phi}&\bigg\{R-  \big[2(1-D)\alpha-2\beta\big]\Box \phi 
		\nonumber\\ 
		&\,- \left[ (D-2)(1-D)\alpha^2+2\beta\big((2-D)\alpha-\beta\big)\right](\partial \phi)^2 \nonumber\\ 
		&\,-\frac{1}{4}e^{2(\beta-\alpha)\phi}F^2\bigg\}.    
	\end{align}
	where $g$ is the determinant of the $D$-dimensional metric. 
	%get rid of the exponential in front of the $D$-dimensional Ricci scalar, that will allow us to 
	A canonical $D$-dimensional Einstein-Hilbert action is obtained for
	\begin{equation}
		\label{relation between alpha and beta}
		(D-2)\alpha+\beta=0.  
	\end{equation}
	and the canonical dilaton kinetic term fixes the constant $\alpha$ to be:
	\begin{equation}
		\label{alpha}
		\alpha^2=\frac{1}{2(D-1)(D-2)}.
	\end{equation}
	Since all fields are independent of $z$, we can perform the integration over this coordinate to obtain, keeping only the zero modes,\footnote{The factor in front of the D'Alambertian operator, $2\alpha$, corrects the expression sometimes found in the literature, $(D-3)\alpha$. As long as only minimal coupling to gravity is considered, the difference is harmless.}
	\begin{equation}
		\mathcal{S}_{0,0}^{(D)}=\frac{2\pi L}{2\hat{\kappa}^2}\int \mathrm{d}^Dx\sqrt{(-1)^{D-1}g}\left[ R+2\alpha \Box \phi+\frac{1}{2}(\partial \phi)^2-\frac{1}{4}e^{2(1-D)\alpha \phi}F^2\right].
	\end{equation}
	We define the $D$-dimensional constant $\kappa$ in terms of the $(D+1)$-dimensional $\hat{\kappa}$ as 
	\begin{equation}
		\label{relation M_P different dimensions}
		\frac{1}{\kappa^2}=\frac{2\pi L}{\hat{\kappa}^2} \Longrightarrow  M_P^{D-2}=2\pi L\,\hat M_P^{D-1}      
	\end{equation}
	In \eqref{metric}, the $\phi$ and $A_{\mu}$ fields  are dimensionless. Dimensional fields, that we denote $\tilde{\phi}$ and $\tilde{A_{\mu}}$, can be written as
	
	\begin{equation}
		\label{physical fields}
		\tilde{\phi}=\frac{\phi}{\sqrt{2}\kappa}; \,\,\,\, \,\,\,\, \tilde{A_{\mu}}=\frac{A_{\mu}}{\sqrt{2}\kappa}
	\end{equation}
	The action of the $D$-dimensional gauge and scalar fields, denoted  as the graviphoton and the dilaton, respectively, reads:
	\begin{equation}
		\label{compactified gravitation action}
		\mathcal{S}_{0,0}^{(D)}=\int \mathrm{d}^Dx\sqrt{(-1)^{D-1}g}\left[ \frac{R}{2\kappa^2}+\frac{\sqrt2\alpha}{\kappa} \Box \tilde{\phi}+\frac{1}{2}(\partial \tilde{\phi})^2-\frac{1}{4}e^{2\sqrt{2}(1-D)\alpha\kappa \tilde{\phi}}\tilde{F}^2\right].
	\end{equation}
	In the following, with the exception of section 5, the second term in  (\ref{compactified gravitation action}), being a total derivative, will be discarded and, for notational simplicity, we remove the tilde in our notation.

	For simplicity, we restrict to the simplest case where the field $\hat\Phi$ is periodic and single-valued on the compact dimension
	\begin{equation}
		\label{scalar decomposition}
		\hat\Phi(x,z+2\pi L)=\hat\Phi(x,z), \qquad \hat\Phi(x,z)=\frac{1}{\sqrt{2\pi L}}\sum_{n=-\infty}^{+\infty}\varphi_n(x)e^{\frac{inz}{L}},
	\end{equation}
	which leads to
	\begin{align}
		\label{action of massless scalar in D dimensions}
		\mathcal{S}=\int \mathrm{d}^Dx\sqrt{(-1)^{D-1}g}\Bigg\{&\frac{R}{2\kappa^2}+\frac{1}{2}(\partial \phi)^2-\frac{1}{4}e^{-2\sqrt\frac{D-1}{D-2}\kappa {\phi}}F^2+\frac{1}{2}\partial_\mu\varphi_0\partial^\mu\varphi_0 \nonumber\\  &+\sum_{n=1}^{\infty}\left(\partial_\mu\varphi_n\partial^\mu\varphi_n^*-\frac{n^2}{L^2}e^{2\sqrt{\frac{D-1}{D-2}}\kappa {\phi}}\varphi_n\varphi_n^*\right) \nonumber \\
		&+\sum_{n=1}^\infty \left(i\sqrt{2}\kappa \frac{ n}{L } A^{\mu}\left(\partial_{\mu}\varphi_n \varphi_n^*-\varphi_n\partial_\mu\varphi_n^*\right)+ {2}\kappa^2 \frac{n^2}{L^2}A_\mu A^\mu\varphi_n\varphi_n^* \right)\Bigg\},
	\end{align}
	where we have chosen in \eqref{alpha} the positive root for $\alpha$. The complex scalars $\varphi_n$ form the Kaluza-Klein (KK) tower and appear minimally coupled to the graviphoton. Around a generic background value $\phi_0$ for the dilaton, the gauge coupling $g$ is given by
	\begin{equation}
		g^2=e^{2\sqrt{\frac{D-1}{D-2}}\kappa {\phi_0}}.  
	\end{equation}
	For each KK mode, the mass and charge read
	\begin{equation}
		\label{charge and mass}
		gq_n=\sqrt{2}\kappa \frac{n}{L}e^{\sqrt{\frac{D-1}{D-2}}\kappa {\phi_0}} \;  \qquad m_n=\frac{n}{L}e^{\sqrt{\frac{D-1}{D-2}}\kappa {\phi_0}}.
	\end{equation}
	This shows that they are related through
	\begin{equation}
		\label{relation mass charge}
		(gq_n)^2=2\kappa^2m_n^2,
	\end{equation}
	saturating the dilatonic WGC condition. This is expected as all the interactions unify to descend from the unique gravitational interaction of a free scalar field in higher dimensions. Useful for the rest of the manuscript is to derive this result proceeding instead with the expansion of the metric \eqref{metric} to second order:
	\begin{equation}
		\hat g_{MN}=\hat\zeta_{MN}+2\hat\kappa\hat h_{MN}+4\hat\kappa^2\hat f_{MN}+o(\hat\kappa^3) 
		\label{metric expansion second order}
	\end{equation}
	where:
	\begin{equation}
		\hat\zeta_{MN}=
		\begin{pmatrix}
			e^{2\sqrt{2}\alpha \hat\kappa{\phi_0}} \eta_{\mu\nu}  & 0 \\
			0 &  -e^{2\sqrt{2}\beta  \hat\kappa{\phi_0}}
		\end{pmatrix}.
	\end{equation}
	is the background metric and $\hat\kappa^2\hat f_{MN}\ll\hat\kappa \hat h_{MN}\ll 1$,   for all $M,N$. We write the perturbation as
	\begin{align}
		\begin{cases}
			&\hat g_{M N}=\hat \zeta_{M N}+2 \kappa \hat h_{M N}+4\kappa^2 \hat f_{M N}+\mathcal O(\kappa^3) \\
			&\hat g^{M N}=\hat \zeta^{M N}+2 \kappa \hat t^{M N}+4\kappa^2 \hat l^{M N}+\mathcal O(\kappa^3).
		\end{cases}
	\end{align}
	The relation $\hat g_{M P}\hat g^{P N}\equiv\delta_{M}^{N}$ reads 
	\begin{equation}
		\begin{cases}
			\hat t^{M N}=-\hat h^{M N} \\
			\hat l^{M N}+ \hat f^{M N}= \hat h^M_P \hat h^{P N},
		\end{cases}
	\end{equation}
	where it is understood that the indices are raised and lowered with the background metric $\hat \zeta$, then
	
	\begin{align}
		\label{decomposition action}
		\sqrt{(-1)^D\hat g}\mathcal L_\Phi
		=& \sqrt{(-1)^D\hat \zeta}\left[\frac{1}{2}\partial_M \hat\Phi\partial^M \hat\Phi-\frac{\hat\kappa'}{2}\hat h^{MN}\left(\partial_M \hat\Phi\partial_N \hat\Phi-\frac{1}{2}\hat\zeta_{MN}\partial_P \hat\Phi\partial^P \hat\Phi\right)  \right. \nonumber \\
		& \left. +\frac{\hat\kappa'^2}{2}\left(\hat l^{MN}-\frac{1}{2}\hat h^{MN}\hat h^P_{\,P}\right)\partial_M\hat\Phi\partial_N\hat\Phi\right.  \nonumber \\
		&\left.+\frac{\hat\kappa'^2}{4}\left(\hat f^P_{\,P}-\frac{1}{2}\hat h_{MP}\hat h^{PM}+\frac{1}{4} (\hat h^P_{\,P})^2\right)\partial_M\hat\Phi\partial^M\hat\Phi\right].
	\end{align}
	where $\hat\kappa'\equiv 2\hat\kappa$. With:
	
	\begin{equation}
		\label{first order h}
		\hat h^{MN}=\frac{1}{\sqrt{2\pi L}}\begin{pmatrix}
			e^{-2\sqrt{2}\alpha \hat\kappa{\phi_0}}\left(
			\sqrt{2}\alpha\phi\,\eta^{\mu\nu}+h^{\mu\nu}\right)
			& -e^{-2\sqrt{2}\alpha  \hat\kappa{\phi_0}}
			\frac{A^\mu}{\sqrt 2}  \\ 
			-e^{-2\sqrt{2}\alpha \hat\kappa{\phi_0}}
			\frac{A^\nu}{\sqrt 2} & -e^{-2\sqrt{2}\beta  \hat\kappa{\phi_0}}
			\sqrt 2\beta\phi
		\end{pmatrix}, 
	\end{equation}
	and using $\sqrt{(-1)^D\hat\zeta}=e^{\sqrt{2}(D\alpha+\beta) \hat\kappa{\phi_0}}$,
	this leads to the coupling between the leading order fluctuations ${\hat h^{MN}}$ of the metric and the stress-energy-momentum of the scalar field $\hat T^{\hat \Phi}_{MN}$:
	\begin{align}
		{\mathcal L}_{int}^{(1)}= -\hat\kappa \hat h^{MN} \hat T^{\hat \Phi}_{MN}=&- \hat\kappa h^{\mu\nu} T^{(\varphi_0,\varphi_n)}_{\mu\nu}\\ &-i \sqrt2 \hat\kappa A^\mu \sum_{n=1}^{\infty}\frac{n}{L}\left(\partial_\mu\varphi_n\,\varphi_n^*-\varphi_n\,\partial_\mu\varphi_n^*\right) -2\sqrt{\frac{D-1}{D-2}} \hat\kappa e^{2\sqrt{\frac{D-1}{D-2}}\hat\kappa \phi_0} \phi \sum_{n=1}^{\infty}\frac{n^2}{L^2}\varphi_n\varphi_n^*. \nonumber
	\end{align}
	Next, we  identify $\hat f_{MN}$ from the metric decomposition at second order:
	\begin{equation}
		\label{second order f}
		\hat{f}_{MN}=\frac{1}{2\pi L}\begin{pmatrix}
			e^{2\sqrt{2}\alpha \hat\kappa{\phi_0}}\bigg(
			\alpha^2\phi^2\eta_{\mu\nu}+\sqrt{2}\alpha\phi\,h_{\mu\nu}+f_{\mu\nu}\bigg)
			-\frac{1}{2}e^{2\sqrt{2}\beta  \hat\kappa{\phi_0}}
			A_\mu A_\nu& e^{2\sqrt{2}\beta  \hat\kappa{\phi_0}}
			\beta\phi A_\mu  \\ 
			e^{2\sqrt{2}\beta  \hat\kappa{\phi_0}}
			\beta\phi A_\nu & -e^{2\sqrt{2}\beta  \hat\kappa{\phi_0}}
			\beta^2\phi^2
		\end{pmatrix}.
	\end{equation}
	With this result, $\hat l^{MN}$ in (\ref{decomposition action}) is given by
	\begin{equation}
		\hat l^{MN}=\hat h^{MP}\hat h_{\,P}^N-\hat f^{MN}  
	\end{equation}
	Using \eqref{second order f} and \eqref{first order h} one obtains
	%% passage comment�
	
	\begin{comment}
		
		\begin{align}
			\hat l^{MN}=\hat h^{MP}\hat h_{\,P}^N-\hat f^{MN}&= 
			\frac{1}{2\pi L}\begin{pmatrix}
				e^{2\sqrt{2}\alpha \hat\kappa{\phi_0}}\left(
				\sqrt{2}\alpha\phi\,\eta^{\mu\rho}+h^{\mu\rho}\right)
				& e^{2\sqrt{2}\beta  \hat\kappa{\phi_0}}
				\frac{A^\mu}{\sqrt 2}  \\ 
				e^{2\sqrt{2}\beta  \hat\kappa{\phi_0}}
				\frac{A^\rho}{\sqrt 2} & e^{2\sqrt{2}\beta  \hat\kappa{\phi_0}}
				\sqrt 2\beta\phi
			\end{pmatrix}\cdot \nonumber\\
			&\cdot\begin{pmatrix}
				e^{2\sqrt{2}\alpha \hat\kappa{\phi_0}}\left(
				\sqrt{2}\alpha\phi\,\eta_\rho^\nu+h_\rho^\nu\right)
				& e^{2\sqrt{2}\beta  \hat\kappa{\phi_0}}
				\frac{A_\rho}{\sqrt 2}  \\ 
				e^{2\sqrt{2}\beta  \hat\kappa{\phi_0}}
				\frac{A^\nu}{\sqrt 2} & e^{2\sqrt{2}\beta  \hat\kappa{\phi_0}}
				\sqrt 2\beta\phi
			\end{pmatrix} \nonumber \\
			&-\frac{1}{2\pi\,L}\begin{pmatrix}
				e^{2\sqrt{2}\alpha \hat\kappa {\phi_0}}\bigg(
				\alpha^2\phi^2\eta_{\mu\nu}+\sqrt{2}\alpha\phi\,h_{\mu\nu}+f_{\mu\nu}\bigg)
				+e^{2\sqrt{2}\beta  \hat\kappa{\phi_0}}
				\frac{A\mu A\nu}{2}& e^{2\sqrt{2}\beta  \hat\kappa{\phi_0}}
				\beta\phi A_\mu  \\ 
				e^{2\sqrt{2}\beta  \hat\kappa{\phi_0}}
				\beta\phi A_\nu & e^{2\sqrt{2}\beta  \hat\kappa{\phi_0}}
				\beta^2\phi^2
			\end{pmatrix},
		\end{align}
		resulting in 
		
	\end{comment}
	%%%%%%%%%%%%%%%%%
	\begin{equation}
		\hat l^{MN} =\frac{1}{2\pi L}\begin{pmatrix}
			e^{-2\sqrt{2}\alpha \hat\kappa {\phi_0}}\bigg(
			\alpha^2\phi^2\eta^{\mu\nu}+\sqrt{2}\alpha\phi\,h^{\mu\nu}+l^{\mu\nu}\bigg)
			& -e^{-2\sqrt{2}\alpha  \hat\kappa{\phi_0}}
			\left(\alpha\phi A^\mu+\frac{1}{\sqrt{2}}h^{\mu\rho}A_\rho\right)  \\ 
			-e^{-2\sqrt{2}\alpha  \hat\kappa{\phi_0}}
			\left(\alpha\phi A^\nu+\frac{1}{\sqrt 2}h_\rho^\nu A^\rho\right) & -e^{-2\sqrt{2}\beta  \hat\kappa{\phi_0}}
			\beta^2\phi^2+e^{-2\sqrt{2}\alpha \hat\kappa{\phi_0}}\frac{1}{2}A_\rho A^\rho
		\end{pmatrix}. 
	\end{equation}
	We define  $J_{\mu,n}=(\varphi_n\partial_{\mu} \varphi_n^*-\varphi_n^*\partial_{\mu} \varphi_n)$, then the second order interaction in the Lagrangian is given by
	\begin{align}
		{\mathcal L}_{int}^{(2)}= \frac{1}{2} \partial_\mu \varphi_0 \partial_\nu \varphi_0&\left[ \left(\frac{f^\rho_{\;\rho}}{2}-\frac{h^{\rho \sigma}h_{\rho \sigma}}{4}+\frac{(h^\rho_{\;\rho})^2}{8}+\frac{1}{2}\left(D^2\alpha^2+2D\beta\alpha+\beta^2-4D\alpha^2-4\beta\alpha+4\alpha^2\right)\phi^2 \right. \right.  \\
		&\left.\left.+\frac{1}{2}((D-2)\alpha+\beta)\phi h^\rho_{\;\rho}\right)\eta^{\mu \nu}+l^{\mu \nu}-\frac{1}{2}h^\rho_{\;\rho}h^{\mu \nu}\right]  \nonumber \\ 
		+\sum_{n=1}^{\infty}\partial_\mu \varphi_n\partial_\nu \varphi_n^*&\left[ \left(\frac{f^\rho_{\;\rho}}{2}-\frac{h^{\rho \sigma}h_{\rho \sigma}}{4}+\frac{(h^\rho_{\;\rho})^2}{8}+\frac{1}{2}\left(D^2\alpha^2+2D\beta\alpha+\beta^2-4D\alpha^2-4\beta\alpha+4\alpha^2\right)\phi^2 \right. \right.\nonumber \\ \nonumber 
		&\left.\left.+\frac{1}{2}((D-2)\alpha+\beta)\phi h^\rho_{\;\rho}\right)\eta^{\mu \nu}+l^{\mu \nu}-\frac{1}{2}h^\rho_{\;\rho}h^{\mu \nu}\right] \nonumber \\ 
		-\sum_{n=1}^{\infty}\frac{n^2}{L^2}|\varphi_n|^2&\left[-A^2e^{\sqrt{2}(D\alpha-\beta)\kappa{\phi_0}}\left(\frac{f^\rho_{\;\rho}}{2}-\frac{h^{\rho \sigma}h_{\rho \sigma}}{4}+\frac{(h^\rho_{\;\rho})^2}{8}+\left(\frac{1}{2}(D\alpha+\beta)-\beta\right)\phi h^\rho_{\;\rho}\right.\right. \nonumber \\
		&\left.\left.+\frac{1}{2}(D^2\alpha^2+2D\alpha\beta+\beta^2-4D\alpha\beta-4\beta^2+4\beta^2)\phi^2\right)\right]\nonumber \\
		-\sum_{n=1}^{\infty}i\frac{n}{L}h^{\rho \sigma}A_{\rho}&J_{\sigma,n}+i\frac{n}{L}A^{\rho}J_{\rho,n}\left(-\frac{h^\sigma_{\;\sigma}}{2}-((D-2)\alpha+\beta)\phi\right) \nonumber
	\end{align}
	This expression simplifies using the relation between $\beta$ and $\alpha$ \eqref{relation between alpha and beta}. In particular, the coefficients of $\phi^2$ and $\phi\, h_{\rho}^{\rho}$ vanish. One obtains 
	\begin{align}
		{\mathcal L}_{int}^{(2)}= \, \, &\frac{1}{2} \partial_\mu \varphi_0 \partial_\nu \varphi_0\left[ \left(\frac{f^\rho_{\;\rho}}{2}-\frac{h^{\rho \sigma}h_{\rho \sigma}}{4}+\frac{(h^\rho_{\;\rho})^2}{8}\right)\eta^{\mu \nu}+l^{\mu \nu}-\frac{1}{2}h^\rho_{\;\rho}h^{\mu \nu}\right]  \nonumber\\ 
		+&\sum_{n=1}^{\infty}\partial_\mu \varphi_n\partial_\nu \varphi_n^*\left[ \left(\frac{f^\rho_{\;\rho}}{2}-\frac{h^{\rho \sigma}h_{\rho \sigma}}{4}+\frac{(h^\rho_{\;\rho})^2}{8}\right)\eta^{\mu \nu}+l^{\mu \nu}-\frac{1}{2}h^\rho_{\;\rho}h^{\mu \nu}\right] \nonumber \\
		-&\sum_{n=1}^{\infty}\frac{n^2}{L^2}|\varphi_n|^2e^{{2\sqrt{2}(D-1)\alpha\kappa\phi_0}}\left(\frac{f^\rho_{\;\rho}}{2}-\frac{h^{\rho \sigma}h_{\rho \sigma}}{4}+\frac{(h^\rho_{\;\rho})^2}{8}\right)  \nonumber \\ \nonumber
		-&\sum_{n=1}^{\infty}\frac{n^2}{L^2}|\varphi_n|^2\left[-A^2+e^{{2\sqrt{2}(D-1)\alpha\kappa\phi_0}}\left(2(D-1)^2\alpha^2\phi^2+(D-1)\alpha\phi h^\rho_{\;\rho}\right)\right]\nonumber \\& +i\frac{n}{L}A^{\rho}\frac{h^\sigma_{\;\sigma}}{2}J_{\rho,n}-i\frac{n}{L}h^{\rho \sigma}A_{\rho}J_{\sigma,n} 
	\end{align}
	which shows how the gauge invariance of the graviphoton is recovered in this expansion at second order in $\hat\kappa$ and exhibits the minimal coupling of the graviphoton to the tower of scalars in $\hat\kappa$.
	%%%%%%%%%%%%%%%%%%%%%%%%%%%%%%%%%%%%%%%%%%%%%%%%%%%%%%%%%%%%%%%%%%%%%%%%%%%%%%%%%%%%%

	%%%%%%%%%%%%%%%%%%%%%%%%%%%%%%%%%%%%%%%%%%%%%%%%%%%%%%%%%%%%%%%%%%%%%%%%%%%%%%%%%%%%%

	%%%%%%%%%%%%%%%%%%%%%%%%%%%%%%%%%%%%%%%%%%%%%%%%%%%%%%%%%%%%%%%%%%%%%%%%%%%%%%%%%%%%%

	%%%%%%%%%%%%%%%%%%%%%%%%%%%%%%%%%%%%%%%%%%%%%%%%%%%%%%%%%%%%%%%%%%%%%%%%%%%%%%%%%%%%%

	\section{Scattering Amplitudes and Weak Gravity Conjectures}
	
	In this section, we will compute diverse $2 \rightarrow 2$ amplitudes in the simple model defined above and compare two sets to be identified, one denoted as {\it gravitational} and the other as  {\it non-gravitational} mediated interactions.\\

	We expand the dilaton around its background value $\phi_0$ as $\phi_0+\phi$ in the action \eqref{action of massless scalar in D dimensions} to obtain:
	\begin{align}
		\label{action without potential}
		\mathcal{S}_f=\int \mathrm{d}^Dx\sqrt{(-1)^{D-1}g}\Bigg\{&\frac{R}{2\kappa^2}+\frac{1}{2}(\partial \phi)^2-\frac{1}{4}e^{-2\sqrt\frac{D-1}{D-2}\kappa \phi_0}\sum_{m=0}^\infty\left(-2\sqrt{\frac{D-1}{D-2}}\kappa \right)^m\frac{\phi^m}{m!}F^2 \nonumber \\
		&+\frac{1}{2}\partial_\mu\varphi_0\partial^\mu\varphi_0 +\sum_{n=1}^{\infty}  \partial_\mu\varphi_n\partial^\mu\varphi_n^*  \nonumber \\
		&-\sum_{n=1}^{\infty}\left(\frac{n^2}{L^2}e^{2\sqrt{\frac{D-1}{D-2}}\kappa{\phi_0}}\sum_{m=0}^\infty\left(2\sqrt{\frac{D-1}{D-2}}\kappa \right)^m\frac{\phi^m}{m!}\varphi_n\varphi_n^*\right) \nonumber \\
		&+\sum_{n=1}^\infty \left(i{\sqrt{2}}\kappa \frac{n}{L}A^{\mu}\left(\partial_{\mu}\varphi_n \varphi_n^*-\varphi_n\partial_\mu\varphi_n^*\right)+{2}\kappa^2 \frac{n^2}{L^2}A_\mu A^\mu\varphi_n\varphi_n^* \right)\Bigg\}
	\end{align}
	where diverse interactions can be identified. For instance:
	
	\begin{itemize}
		\item  3 and 4-point vertices for minimally-coupled scalars to graviphotons appear in the last line. We can identify the KK electric charges 
		\begin{equation}
			gq_n={\sqrt 2}\kappa \frac{n}{L} e^{\sqrt{\frac{D-1}{D-2}}\kappa\,  {\phi_0}}.   
		\end{equation}
		\\
		
		\item In the third line, the $m$-th term ($m\neq 0$) in the sum gives a $(2+m)$-point interaction with $m$ dilatons and two KK scalars with coupling 
		\begin{equation}
			-i\left(2\sqrt{\frac{D-1}{D-2}}\kappa \right)^m\, \frac{n^2}{L^2} \,  e^{2\sqrt{\frac{D-1}{D-2}}\kappa {\phi_0}}.
		\end{equation}
		\\
		
		\item The $m$-th term in the sum in front of $F^2$ in the first line gives a coupling of $m$ dilatons with two gauge fields 
		\begin{equation}
			-i\left(-2\sqrt{\frac{D-1}{D-2}}\kappa \right)^m \left(p_1\cdot p_2 \,\eta_{\mu\nu}-p_{1\,\nu}p_{2\,\mu}\right).
		\end{equation} 
	\end{itemize}
	Expansion of the metric around flat space-time $g_{\mu\nu}=\eta_{\mu\nu}+2\kappa h_{\mu\nu}$ gives the usual minimal couplings to gravity for both the matter fields ($\varphi_0$, $\varphi_n$) and the massless mediators ($\phi$, $A_\mu$).

	%%%%%%%%%%%%%%%%%%%%%%%%%%%%%%%%%%%%%%%%%%%%%%%%%%%%%%%%%%%%%%%%%%%%%%%%%%%%%%%%%%%%%

	%%%%%%%%%%%%%%%%%%%%%%%%%%%%%%%%%%%%%%%%%%%%%%%%%%%%%%%%%%%%%%%%%%%%%%%%%%%%%%%%%%%%%

	%%%%%%%%%%%%%%%%%%%%%%%%%%%%%%%%%%%%%%%%%%%%%%%%%%%%%%%%%%%%%%%%%%%%%%%%%%%%%%%%%%%%%

	\subsection{The Dilatonic WGC}
	\label{SectionforcebetweenKKstates}
	
	Consider the tree-level $2\to 2$ scattering\footnote{We adopt here this simple notation where $\varphi_n(p)$, or $\ket{\varphi_n(p)}$ should not be viewed as the field operator acting on the vacuum but to represent a one-particle state of momentum $p$.}$^,$\footnote{Here and throughout, $s, t$ and $u$ will denote the Mandelstam variables.} $\varphi_n(p_1)\varphi_n(p_2) \rightarrow \varphi_n(p_3) \varphi_n(p_4)$:
	
	\begin{align}
		i\mathcal M=&ig^2q_n^2\left(\frac{(p_1+p_3)\cdot(p_2+p_4)}{t}+\frac{(p_1+p_4)\cdot(p_2+p_3)}{u}\right)  -4i\frac{D-1}{D-2}\kappa^2 m^4_n\left(\frac{1}{t}+\frac{1}{u}\right) \nonumber \\
		&-\frac{\kappa^2}{4}\Bigg[\Big(p_{1\mu}p_{3\nu}+p_{3\mu}p_{1\nu}-\eta_{\mu\nu}\left(p_1\cdot p_3-m_n^2\right)\Big) \frac{i\mathcal P^{\mu\nu\alpha\beta}}{t}\Big(p_{2\alpha}p_{4\beta}+p_{4\alpha}p_{2\beta}-\eta_{\alpha\beta}\left(p_2\cdot p_4-m_n^2\right)\Big)\nonumber \\
		&\qquad\quad+(t,p_3,p_4)\leftrightarrow (u,p_4,p_3) \Bigg]
	\end{align}
	where $\mathcal{P}$ is the usual massless spin-2 projector 
	\begin{equation}
		\mathcal P^{\alpha\beta \rho\sigma}=\frac{\eta^{\alpha\rho}\eta^{\beta\sigma}+\eta^{\alpha\sigma}\eta^{\beta\rho}}{2}-\frac{\eta^{\alpha\beta}\eta^{\rho\sigma}}{D-2}
	\end{equation} 
	and we have separated the contributions from the exchanges of the gauge boson, the dilaton and the graviton, respectively. \\
	Taking the non-relativistic (NR) limit 
	\begin{equation}
		\frac{s-4m_n^2}{m_n^2}\to 0,\qquad \frac{t}{m_n^2}\to 0,\qquad {\rm and} \quad \frac{u}{m_n^2} \to 0
	\end{equation}
	%${s-4m_n^2}/{m_n^2},\qquad {t}/{m_n^2},\qquad {u}/{m_n^2} \to 0$
	and expressing the charge in terms of the mass we obtain 
	\begin{equation}
		i\mathcal M \to i\mathcal M_{NR}=4im_n^2\left[g^2q_n^2-\kappa^2m_n^2\left(\frac{D-1}{D-2}+\frac{D-3}{D-2}\right)\right]\left(\frac{1}{t}+\frac{1}{u}\right)=0.
	\end{equation}
	The relation between the charge and the mass \eqref{relation mass charge}  ensures the cancellation between the three forces.  \\
	It is straightforward to generalize this to see that dominance of the gauge interaction requires that a state with charge $q$ and mass $m$ satisfying the relation 
	\begin{equation}
		\label{generic dilatonic WGC}
		g^2q^2\ge \left(\frac{\alpha^2}{2}+\frac{D-3}{D-2}\right)\kappa^2 m^2,
	\end{equation}
	where $\alpha$ is the dilatonic coupling of the form $e^{2\sqrt{2}\alpha \kappa\phi}F^2$, exists. We have therefore recovered in this explicit amplitude computation the Dilatonic Weak Gravity Conjecture that was derived in \cite{Heidenreich:2015nta} (see also \cite{Benakli:2021fvv} for its generalization) from the study of the extremal Einstein-Maxwell-dilaton black hole solutions.  In the absence of the massless dilaton field $\alpha=0$, one trivially retrieves the original WGC condition 
	\begin{equation}
		\label{WGC-Dimension}  g^2q^2\ge\frac{D-3}{D-2}\kappa^2 m^2.   
	\end{equation}
	
	%%%%%%%%%%%%%%%%%%%%%%%%%%%%%%%%%%%%%%%%%%%%%%%%%%%%%%%%%%%%%%

	%%%%%%%%%%%%%%%%%%%%%%%%%%%%%%%%%%%%%%%%%%%%%%%%%%%%%%%%%%%%%%

	\subsection{Amplitudes for Pair Production }
	\label{pair production subsection}

	Consider the production of a pair of matter states, here scalar KK states, of momenta $p_3,\, p_4$ from {\it massless} particles of momenta $p_1,\,p_2$. We can split the production processes into two sets:
	\begin{itemize}
		\item Non-gravitational production: a pair of KK scalar modes $\ket{\varphi_n, \varphi_n^*}$ can arise from a pair of photons $\bra{\gamma, \gamma}$, a pair of dilatons $\bra{\phi,\phi}$, or a dilaton and a photon $\bra{\phi , \gamma}$.
		
		\item Gravitational production: this includes the presence of a graviton $G$ in initial states as $\bra{G, G}$, $\bra{G, \gamma}$ or $\bra{G, \phi}$, but also gravitons as intermediate states in the production from $\bra{\gamma, \gamma}$ or $\bra{\phi, \phi}$.  For later convenience, we further divide the gravitational production processes into purely gravitational (the $\bra{G,G}$ production) and mixed (all the others). 
	\end{itemize}
	
	\subsubsection{Non gravitational amplitudes}
	\begin{figure}
		\centering
		\includegraphics[scale=0.75]{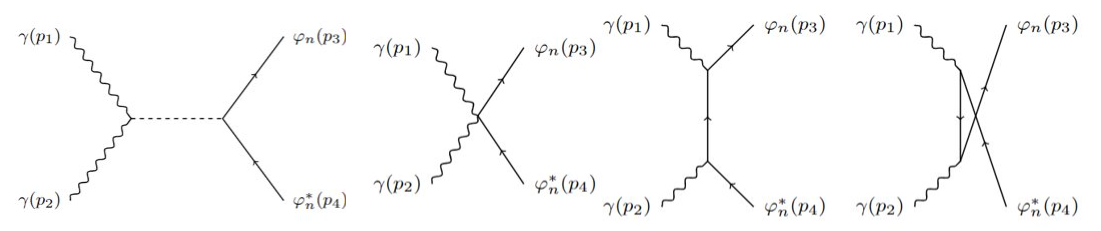}
		\includegraphics[scale=0.75]{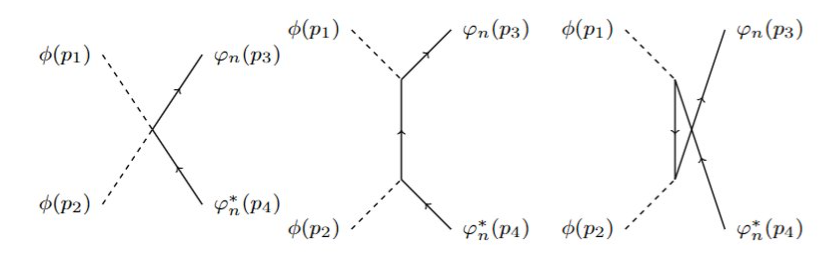}
		\includegraphics[scale=0.75]{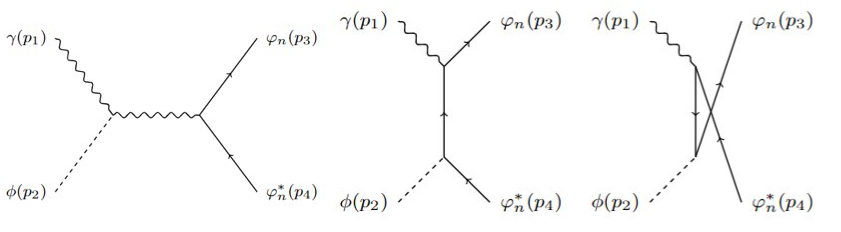}
		\caption{Feynman diagrams for the non-gravitational production of a pair of matter states $\varphi_n,\varphi_n^*$ from two photons (first line), two dilatons (second line) and a dilaton and a photon (third line).}
		\label{non gravitational pair production figure}
	\end{figure}
	
	The production from photons $\gamma \gamma \rightarrow \varphi_n \varphi_n^*$ occurs through the coupling to the $U(1)$ gauge boson plus an s-channel term mediated by the dilaton, as depicted in the first line of figure \ref{non gravitational pair production figure}. These give:
	\begin{align}
		\label{amplitude photon photon pair production}
		i\mathcal M_{\gamma\gamma}=&ig^2q_n^2\, \,   \epsilon_\mu(p_1)\epsilon_\nu(p_2)\left(\frac{(2 p_3^\mu-p_1^\mu) (2 p_4^\nu-p_2^\nu)}{t-m_n^2}+\frac{(2 p_4^\mu-p_1^\mu)(2 p_3^\nu-p_2^\nu)}{u-m_n^2}+2\eta^{\mu\nu}\right) \\ \nonumber &-2i g^2q_n^2\frac{D-1}{D-2}\, \, \, \epsilon_\mu(p_1)\epsilon_\nu(p_2)\frac{p_1\cdot p_2 \eta^{\mu\nu}-p_1^\nu p_2^\mu}{s}.
	\end{align}
	We are interested in the threshold limit
	\begin{equation}
		\frac{s-4m_n^2}{m_n^2}\to 0,\qquad \frac{t+m_n^2}{m_n ^2}\to 0,\qquad \frac{u+m_n^2}{m_n^2}\to 0,   
	\end{equation}  
	leading to 
	\begin{align}
		\label{photon photon pair production NR limit}
		\left|\mathcal M_{\gamma\gamma}\right|^2  \xrightarrow[\mathrm{Threshold}] {}  & \frac{4}{(D-2)^2}\left[(D-2)  -\frac{3}{4}\frac{(D-1)^2}{(D-2)}
		+\frac{D-1}{D-2}\right]g^4q_n^4=\left(\frac{D-3}{D-2}\right)^2\frac{g^4q_n^4}{D-2}
	\end{align}
	We note that, in a $U(1)$ gauge theory with no dilaton, the amplitude would be given by the first line of \eqref{amplitude photon photon pair production} only, that means in the threshold limit $4 g^4q^4/(D-2) $ for a state of charge $q$. 
	
	The  production from a dilation pair $\phi \phi \rightarrow \varphi_n \varphi_n^*$ (second line of figure \ref{non gravitational pair production figure}) is immediately recognized to give a null result in the limit of interest:
	\begin{align}
		i\mathcal M_{\phi\phi}= & - 4i\kappa^2\frac{D-1}{D-2}m_n^4\left(\frac{1}{t-m_n^2}+\frac{1}{u-m_n^2}\right)-4i\kappa^2\frac{D-1}{D-2}m_n^2 \qquad 
		\xrightarrow[\mathrm{Threshold}] {} 0.
	\end{align}
	
	Finally, the production  from the pair photon-dilaton $\phi \gamma \rightarrow \varphi_n \varphi_n^*$ receives contributions from the three $s,t\, \mathrm{and}\, u$-channels (see the third line of figure \ref{non gravitational pair production figure})
	\begin{align}
		\label{amplitude photon scalar pair production}
		i\mathcal M_{\gamma(p_1)\phi(p_2)}&=\epsilon_\mu(p_1) \Bigg\{-2\sqrt{\frac{D-1}{D-2}}\kappa gq_n\left(p_1\cdot(p_1+p_2)g^{\mu\rho}-p_1^\rho (p_1+p_2)^\mu\right)(p_3-p_4)_{\rho}\frac{i}{s} \nonumber\\
		&\hspace{2cm}+2i\sqrt{\frac{D-1}{D-2}}\kappa gq_nm_n^2\left(\frac{(2p_3-p_1)^\mu}{t-m_n^2}-\frac{(2p_4-p_1)^\mu}{u-m_n^2}\right)\Bigg\}, 
	\end{align}
	%We then obtain for $\left|\mathcal M_{\gamma\phi}\right|^2$:
	\begin{comment}
		thus
		\begin{align}
			\left|\mathcal M_{\gamma\phi}\right|^2= 4\frac{D-1}{D-2}\frac{g^2q_n^2}{M_P^{(D-2)/2}}&\Bigg\{(p_1\cdot p_2 g^{\mu\rho}-p_1^\rho p_2^\mu)(p_1\cdot p_2g_\mu^\sigma-p_1^\sigma p_{2\,\mu})\frac{(p_{3\,\rho}-p_{4\,\rho})(p_{3\,\sigma}-p_{4\,\sigma})}{s^2}\nonumber \\
			&+4m_n^4\left(\frac{p_3^\mu}{t-m_n^2}-\frac{p_4^\mu}{u-m_n^2}\right)\left(\frac{p_{3\,\mu}}{t-m_n^2}-\frac{p_{4\,\mu}}{u-m_n^2}\right)+ \nonumber \\
			&-4m_n^2(p_1\cdot p_2 g^{\mu\rho}-p_1^\rho p_2^\mu)\frac{p_{3\,\rho}-p_{4\,\rho}}{s}\left(\frac{p_{3\,\mu}}{t-m_n^2}-\frac{p_{4\,\mu}}{u-m_n^2}\right)\Bigg\}
		\end{align}
	\end{comment}
	and this is easily verified to give a null contribution in the threshold limit.

	\subsubsection{Mixed amplitudes}
	\label{subsection mixed amplitudes}
	
	We consider now the ``mixed gravitational" processes: we start by computing the graviton s-channel mediation for $\gamma\gamma$ and $\phi\phi$ initial states, then the amplitudes with initial states $\gamma\, G$ and $\phi \, G$. We present hereafter the results for the particular case $D=4$. When it will be of interest, we will show the results for a generic number of dimensions $D$. \\
	
	\begin{figure}
		\centering
		\includegraphics[scale=0.8]{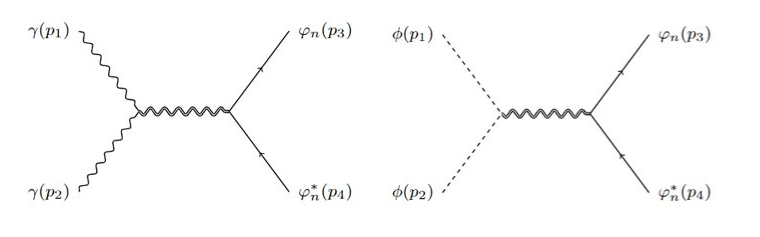}
		\caption{Feynman diagrams for pair production, gravitationally mediated, from photons and dilatons.}
		\label{Gravitationally mediated pair prod figure}
	\end{figure}
	
	The additional contribution to the $\gamma \gamma$ and $\phi\phi$ productions described in figure \ref{Gravitationally mediated pair prod figure} respectively read
	\begin{align}
		i\mathcal M_{\gamma\gamma}^{\mathrm{G}}=-\kappa^2 \Big\{&(p_1\cdot p_2)(\epsilon_{1\,\alpha}\epsilon_{2\,\beta}+\epsilon_{1\,\beta}\epsilon_{2\,\alpha})+(p_{1\,\alpha}p_{2\,\beta}+p_{1\,\beta}p_{2\,\alpha})(\epsilon_1\cdot \epsilon_2)-(\epsilon_{1\,\alpha}p_{2\,\beta}+p_{2\,\alpha}\epsilon_{1\,\beta})(p_1\cdot \epsilon_2) \nonumber  \\
		&-(\epsilon_{2\,\alpha}p_{1\,\beta}+p_{1\,\alpha}\epsilon_{2\,\beta})(p_2\cdot \epsilon_1)-\eta_{\alpha \beta}(p_1\cdot p_2\, \epsilon_1\cdot \epsilon_2-\epsilon_1\cdot p_2\,\epsilon_2\cdot p_1)\Big\}\frac{i\mathcal P^{\alpha\beta\rho\sigma}}{s}\nonumber \\
		&\Big\{p_{3\,\rho}p_{4\,\sigma}+p_{3\,\sigma}p_{4\,\rho}-\eta_{\rho\sigma}(p_3\cdot p_4+m^2)\Big\},
	\end{align}
	and  
	\begin{align}
		i\mathcal M_{\phi\phi}^{\mathrm{G}}= -\kappa^2 \Big\{p_{1\,\alpha}p_{2\,\beta}+p_{1\,\beta}p_{2\,\alpha}-\eta_{\alpha\beta}p_1\cdot p_2\Big\}\frac{i\mathcal P^{\alpha\beta\rho\sigma}}{s}\Big\{p_{3 \,\rho}p_{4 \,\sigma}+p_{3 \,\sigma}p_{4\, \rho}-\eta_{\rho\sigma}(p_3\cdot p_4+m^2)\Big\},
	\end{align}
	where $\epsilon_{i}=\epsilon(p_i)$.
	For the $\gamma\gamma \to \varphi_n \varphi_n^*$ amplitude, a (simpler) way to compute this is through projecting onto a specific basis for the polarizations $\epsilon$ (see Appendix \ref{Helicity method appendix}).

	Working in the center of mass frame for the massive particles, we obtain the different components of the graviton mediated $\gamma \gamma \to \varphi_n \varphi_n^*$ as follows
	\begin{align}
		&i\mathcal M^{\mathrm{G}}_{+,+}=i\mathcal M^{\mathrm{G}}_{-,-}=-i\frac{\kappa^2}{s}\left[tu-m_n^4+(m_n^2-u)^2+su%-\frac{D-4}{D-2}m_n^2 s
		\right]=%i\kappa^2\frac{D-4}{D-2}m_n^2 
		0\nonumber \\
		&i\mathcal M^{\mathrm{G}}_{+,-}=i\mathcal M^{\mathrm{G}}_{-,+}=i\frac{\kappa^2}{s}\left[tu-m_n^4 \right],
	\end{align}
	where the $\pm$ sign refers to the helicities of the incoming gauge bosons.
	In the threshold limit the graviton mediated contribution vanishes for both components. 
	
	In $D$ dimensions, the whole $\mathcal M_{\gamma\gamma}$ amplitude reads
	
	\begin{equation}
		\left|\mathcal M_{\gamma\gamma}\right|^2 \xrightarrow[\mathrm{Threshold}] {}\frac{\left(2 (D-2) (gq)^2+(D-4) \kappa ^2 m^2\right)^2}{(D-2)^3}
	\end{equation}
	for a generic $U(1)$ gauge theory (i.e. when the dilaton is put to zero) and 
	\begin{equation}
		\left|\mathcal M_{\gamma\gamma}\right|^2 \xrightarrow[\mathrm{Threshold}] {}\frac{\left((D-3) (gq_n)^2+(D-4) \kappa ^2 m_n^2\right)^2}{(D-2)^3}
	\end{equation}
	in the dilatonic theory we are studying here. Both the results for the $U(1)$ and dilatonic theory (\eqref{photon photon pair production NR limit} and discussion below) are recovered in the limit $\kappa\to 0$. It is instructive to note, from these equations, that the vanishing of the graviton mediated contribution to the production from a photon pair is specific to the case of $D=4$ dimensions, and in $D\ne 4$ dimensions mixed terms of the form $g^2q^2\times \kappa^2m^2$ are generated.

	For the $\phi\phi \to  \varphi_n \varphi_n^*$ the amplitude reads
	\begin{equation}
		i\mathcal M_{\phi\phi}^{\mathrm{G}}=-i\frac{\kappa^2}{s}\left[m_n^4-ut-m_n^2 s \right].
	\end{equation}
	This results in a non vanishing contribution in the limit of interest such that 
	\begin{equation}\label{scalar gravity mediated}
		i\mathcal M_{\phi\phi}^{\mathrm{G}} = i\kappa^2m_n^2.
	\end{equation}
	
	\begin{figure}
		\centering
		\includegraphics[scale=0.8]{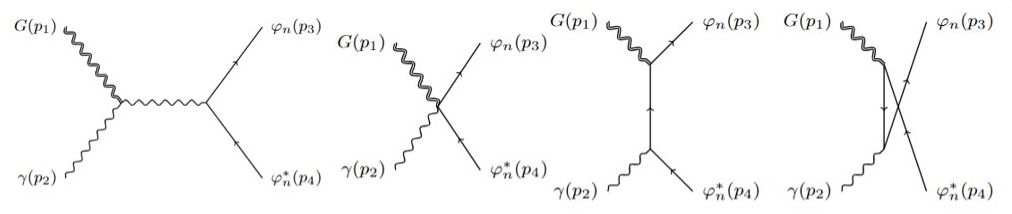}
		\caption{Feynman diagrams for the mixed pair production from a graviton and a photon.}
		\label{Graviton photon pair production figure}
	\end{figure}
	
\begin{figure}
		\centering
		\includegraphics[scale=0.8]{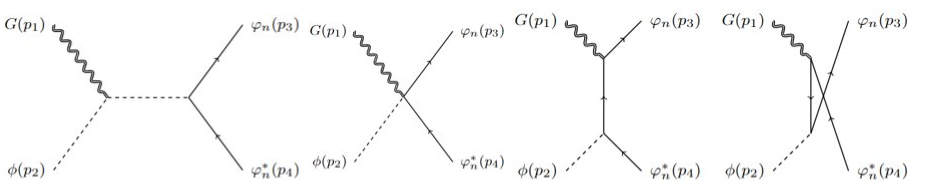}
		\caption{Feynman diagrams for the mixed pair production from a graviton and a dilaton.}
		\label{Graviton dilaton pair production figure}
	\end{figure}

	Concerning the mixed initial states, we have both $\gamma\, G\to  \varphi_n \varphi_n^*$ (see figure \ref{Graviton photon pair production figure}) and $\phi\, G\to \varphi_n \varphi_n^*$ (see figure \ref{Graviton dilaton pair production figure}). Each of these two processes receive contributions from four diagrams. \\ 
	Starting with the graviton-photon production, the amplitude $G(p_1)\gamma(p_2)\to  \varphi_n \varphi_n^*$ takes the form 
	
	\begin{align}
		\label{graviton photon production}
		i\mathcal M^{\mathrm{mix.}}_{G\gamma}=i\kappa gq_n\Bigg(&\frac{4(\epsilon_1\cdot p_3)^2\epsilon_2\cdot p_4}{t-m_n^2}-\frac{4(\epsilon_1\cdot p_4)^2\epsilon_2\cdot p_3}{u-m_n^2}+2\epsilon_1\cdot \epsilon_2 \epsilon_1\cdot(p_3-p_4)\nonumber \\
		&-\frac{(p_1+p_2)\cdot p_2(2\epsilon_1\cdot \epsilon_2 \epsilon_1\cdot(p_3-p_4)}{s}\Bigg)
	\end{align}
	and so for the different choices of graviton and photon helicities:
	
	\begin{equation}
		\begin{cases}
			i\mathcal M^{\mathrm{mix.}}_{++,+}=- i\mathcal M^{\mathrm{mix.}}_{--,-}=-i\kappa gq_n\sqrt{2\frac{tu-m_n^4}{s}}\left(\frac{m_n^4-tu}{(t-m_n^2)(u-m_n^2)}+3\right) \\
			i\mathcal M^{\mathrm{mix.}}_{++,-}=- i\mathcal M^{\mathrm{mix.}}_{--,+}=i\kappa gq_n\sqrt{2\frac{tu-m_n^4}{s}}\left(\frac{m_n^4-tu}{(t-m_n^2)(u-m_n^2)}\right). 
		\end{cases}
	\end{equation}
	It is immediately verified that all these contributions vanish in the threshold limit where $t\to -m_n^2$ and $u\to -m_n^2$. 
	
	The same vanishing limit at threshold holds for the mixed graviton-dilaton production, where the amplitude is 
	\begin{equation}
		\label{graviton dilaton production}
		i\mathcal M^{\mathrm{mix.}}_{G\phi}=-2i\kappa\mu_n\left(\frac{(\epsilon_1\cdot p_3)^2}{t-m_n^2}+\frac{(\epsilon_1\cdot p_4)^2}{u-m_n^2}\right)
	\end{equation}
	with $\mu_n=\sqrt{6}\kappa m_n^2$ the three-point $\phi\varphi_n\varphi_n^*$ $D=4$ coupling, and finally
	\begin{equation}
		i\mathcal M^{\mathrm{mix.}}_{++}= i\mathcal M^{\mathrm{mix.}}_{--}=i\kappa \mu_n \frac{tu-m_n^4}{(t-m_n^2)(u-m_n^2)}.
	\end{equation}
	From the explicit results presented in Appendix \ref{Helicity method appendix}, it is also immediate to realize that the mixed contributions vanish at threshold for all $D$.

	\subsubsection{Gravitational production amplitudes}
	\label{subsection purely gravitational pair production}

\begin{figure}
		\centering
		\includegraphics[scale=0.8]{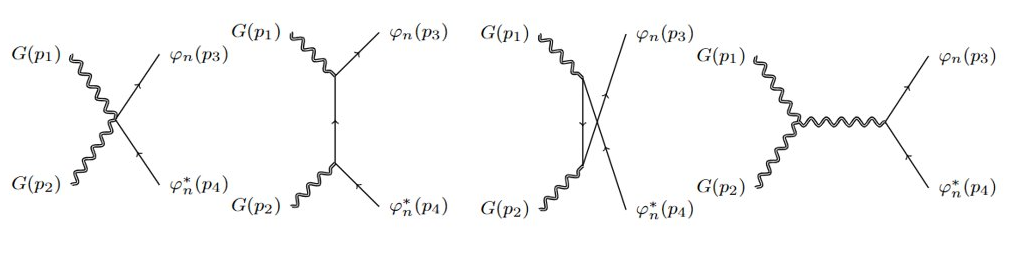}
		\caption{Feynman diagrams for the production of a pair of matter states from two gravitons.}
		\label{graviton pair production figure}
	\end{figure}

	Finally, we discuss the purely gravitational production. The starting point for the expression of the amplitude is rather long. It receives in fact contribution from the four diagrams of figure \ref{graviton pair production figure}, each one with vertices determined from a two-derivative interacting term (some details about two-derivative interactions are discussed in Appendix \ref{Appendix A}). We prefer to give here a more compact expression that is obtained after some algebra:
	
	\begin{align}\label{graviton graviton amplitude}
		i\mathcal M_{GG}=&\frac{\kappa ^2}{2} \left(-\frac{8 (p_3\cdot\epsilon_1)^2 (p_4\cdot\epsilon_2)^2}{t-m_n^2}-\frac{8 (p_3\cdot\epsilon_2)^2 (p_4\cdot\epsilon _1)^2}{u-m_n^2} \right.\nonumber\\
		&\left.\qquad-2\frac{(\epsilon_1\cdot\epsilon _2)^2 \left(m_n^4-tu-sm_n^2\right)}{s}-4 \epsilon_1\cdot\epsilon_2 \left(p_3\cdot\epsilon_2\, p_4\cdot\epsilon_1+p_3\cdot\epsilon_1\, p_4\cdot\epsilon_2\right)\right)
	\end{align}
	The complete results for each one of the four diagrams contributing to the amplitude are presented in Appendix \ref{Helicity method appendix}, together with the description of the helicity method.
	Using now the specific basis for $D=4$ dimensions, we find
	\begin{align}
		i\mathcal M_{++,++}=i\mathcal M_{--,--}&=i\kappa ^2  \left(\frac{\left(m_n^4-t u\right)m^2}{(t-m_n^2)(u-m_n^2)}+ m_n^2\right)  \nonumber\\
		i\mathcal M_{++,--}=i\mathcal M_{--,++}&=i\kappa ^2 \frac{\left(m_n^4-t u\right)^2}{s \left(t-m_n^2\right) \left(u-m_n^2\right)},
	\end{align}
	Comparing this result with the one obtained from the $\gamma\gamma$ production in the case with no dilaton, we verify the factorization
	
	\begin{align}\label{factorization}
		\mathcal M^{(GG)}_{++,++}=&\frac{\kappa ^2}{4 (gq)^4}\frac{\left(t-m_n^2\right) \left(u-m_n^2\right)}{s} \mathcal M^{(\gamma\gamma)}_{+,+}\nonumber\\
		\mathcal M^{(GG)}_{++,--}=&\frac{\kappa ^2}{4 (gq)^4}\frac{\left(t-m_n^2\right) \left(u-m_n^2\right)}{s} \mathcal M^{(\gamma\gamma)}_{+,-}.
	\end{align} 
	The corresponding factorization for the comparison between the gravitational Compton scattering $G\varphi\to G\varphi$ (with $\varphi$ a generic scalar field) and the usual Compton scattering was found in \cite{Choi:1993xa,Choi:1993wu} (see also \cite{Holstein:2006bh}).
	
	From the above results, in the threshold limit we have 
	
	\begin{equation}
		\left|\mathcal M_{GG}\right|^2=\frac{1}{4}\left(\left|\mathcal M_{++,++}\right|^2+\left|\mathcal M_{++,--}\right|^2+\left|\mathcal M_{--,++}\right|^2+\left|\mathcal M_{--,--}\right|^2\right)\to \frac{\kappa^4 m_n^4}{2}.
	\end{equation}
	Note that the result $ \left|\mathcal M_{GG}\right|^2\to \kappa^4m^4/2$, and more generally the "purely gravitational" pair production, is independent from the presence of the dilaton. 
	This is easily generalized to the case of generic $D$ (see again Appendix \ref{Helicity method appendix} for details) and leads in the threshold limit to
	\begin{equation}
		\left|\mathcal M_{GG}\right|^2\to \frac{1}{D-2}\kappa^4 m_n^4  .
	\end{equation}

	\subsubsection{Gravitational vs gauge amplitudes}

	When the dilaton is put to zero, the requirement 
	\begin{equation}
		\label{pair production comparison}
		\left|\mathcal M_{\gamma\gamma}\right|^2\underset{\mathrm{Threshold}}{\ge}\left|\mathcal M_{GG}\right|^2
	\end{equation}
	gives the original $U(1),\, D=4$ WGC bound $\sqrt2 gq\ge \kappa m$. 
	
	Using cross-symmetry on the results of \cite{Choi:1993xa,Choi:1993wu,Holstein:2006bh}, the authors of \cite{Gonzalo:2020kke} observed that \eqref{pair production comparison} leads to the WGC relation and proposed \eqref{pair production comparison} as a possible alternative formulation of the WGC. In \cite{Gonzalo:2020kke}, the graviton-mediated diagram was not taken into account in the $\gamma$ amplitude. Our calculation shows that in the threshold limit, the contribution of this additional diagram disappears. Therefore, in the four-dimensional $U(1)$ gauge theory, we can safely compare, as in the \eqref{pair production comparison}, the $\gamma\gamma$ and $GG$ productions without having to neglect any contribution. 
	
	Our calculation also shows that in $D=4$ dimensions, the KK states saturate \eqref{pair production comparison}. In fact, we emphasize again that the gravitational amplitude $\mathcal M_{GG}$, here, does not care about the presence of the dilaton: whether the theory is a simple $U(1)$ or a dilatonic $U(1)$, the result for $\mathcal M_{GG}$ is unchanged. On the other hand, the amplitude $\mathcal M_{\gamma\gamma}$ receives an additional contribution which changes the numerical coefficient in front of $g^4q^4$ from $2$ to $1/8$. Since the $\mathcal M_{\gamma\phi}$ and $\mathcal M_{\phi\phi}$ amplitudes both vanish in the threshold limit, the comparison of the pair production processes in this KK theory leads to
	\begin{equation}
		\frac{g^4 q_n^4}{8}\ge \frac{\kappa^4 m_n^4}{2}\Longrightarrow gq\ge\sqrt{2}\kappa m
	\end{equation}
	and \eqref{relation mass charge} shows that KK states saturate it.  
	
	However, if, in the presence of the dilaton, we consider gravitationally mediated diagrams for $\gamma \gamma$ and $\phi \phi$ amplitudes, there is a non-vanishing contribution that comes from $\mathcal M_{\phi\phi}^{\mathrm G}$ in \eqref{scalar gravity mediated}, and this would clearly spoil the saturation observed for the KK states. The inclusion of the mixed production channels $G\gamma$\,\eqref{graviton photon production} and $G\phi$\,\eqref{graviton dilaton production} cannot restore the saturation property, since both do not contribute in the limit of interest. The dilatonic WGC will be recovered only if the contributions from graviton exchanges in $\gamma\gamma$ and $\phi \phi$ amplitudes are not included.
	
	Note also that the pairwise production comparison does not reproduce the constraints of WGCs in more than 4 dimensions.  The $\mathcal{M}_{\gamma\gamma}$ and $\mathcal{M}_{GG}$ amplitudes lead, for any $D$, to compare $\sqrt{2} gq$ and $\kappa m$. For the case of a simple theory $U(1)$, setting as quoted above the dilaton to zero in our calculations, the result for the production from a photon pair in $D$ dimensions in the threshold limit is
	\begin{equation}\label{d dimensions gauge}
		\left|\mathcal M_{\gamma\gamma}\right|^2=\frac{4}{D-2}(gq)^4. 
	\end{equation}
	In Appendix \ref{Helicity method appendix} we learn that the purely gravitational production of pairs gives, in the same limit of interest,
	\begin{equation}\label{d dimensions gravity}
		\left|\mathcal M_{GG}\right|^2=\frac{1}{D-2}(\kappa m)^4.
	\end{equation}
	By comparing \eqref{d dimensions gauge} and  \eqref{d dimensions gravity}, it is immediate to observe that requiring $\left|\mathcal M_{\gamma}\right|^2 \ge \left|\mathcal M_{GG}\right|^2$, one {\it does not reproduce} the WGC bound
	\begin{equation}
		gq\ge\sqrt{\frac{D-3}{D-2}}\kappa m.
	\end{equation}
	
	Similarly, the comparison of purely gravitational pair production and purely non-gravitational pair production in the KK theory we consider here amounts to a comparison of the results
	\begin{equation}
		\left|\mathcal M_{\gamma\gamma}\right|^2\to\frac{(D-3)^2}{(D-2)^3}g^4q_n^4,\qquad \left|\mathcal M_{\phi\phi}\right|^2\to 0,\qquad \left|\mathcal M_{GG}\right|^2 \to \frac{1}{D-2}\kappa^4 m_n^4.
	\end{equation}
	Using \eqref{relation mass charge}, it is immediate to realize that the KK states saturate the \eqref{pair production comparison} (or an equivalent generalization of it to include the $\mathcal M_{\phi\phi}$ contribution which disappears here) only for $D=4$. The results of section \ref{subsection mixed amplitudes} show that the addition of mixed contributions does not change this. 
	%%%%%%%%%%%%%%%%%%%%%%%%%%%%%%%%%%%%%%%%%%%%%%%%%%%%%%%%%%%%%%%%%%%%%%%%%%%%%%%%%%%%%

	%%%%%%%%%%%%%%%%%%%%%%%%%%%%%%%%%%%%%%%%%%%%%%%%%%%%%%%%%%%%%%%%%%%%%%%%%%%%%%%%%%%%%

	%%%%%%%%%%%%%%%%%%%%%%%%%%%%%%%%%%%%%%%%%%%%%%%%%%%%%%%%%%%%%%%%%%%%%%%%%%%%%%%%%%%%%

	%%%%%%%%%%%%%%%%%%%%%%%%%%%%%%%%%%%%%%%%%%%%%%%%%%%%%%%%%%%%%%%%%%%%%%%%%%%%%%%%%%%%%

	%%%%%%%%%%%%%%%%%%%%%%%%%%%%%%%%%%%%%%%%%%%%%%%%%%%%%%%%%%%%%%%%%%%%%%%%%%%%%%%%%%%%%

	%%%%%%%%%%%%%%%%%%%%%%%%%%%%%%%%%%%%%%%%%%%%%%%%%%%%%%%%%%%%%%%%%%%%%%%%%%%%%%%%%%%%%

	\section{Massive and Self-interacting Scalars}

	%%%%%%%%%%%%%%%%%%%%%%%%%%%%%%%%%%%%%%%%%%%%%%%%%%%%%%%%%%%%%%%%%%%%%%%%%%%%%%%%%%%%%

	%%%%%%%%%%%%%%%%%%%%%%%%%%%%%%%%%%%%%%%%%%%%%%%%%%%%%%%%%%%%%%%%%%%%%%%%%%%%%%%%%%%%%

	We next consider the presence of mass and self-interacting terms in the higher dimensional scalar theory. The KK scalar modes are no more extremal states of the WGC, but this set-up will allow us to retrieve Scalar Weak Gravity Conjectures which are postulated to constrain the relative strength of the additional terms.

	We will consider the simple extension of \eqref{free scalar action}
	
	\begin{equation}
		\label{actioninteractions}
		\mathcal S_{int}=\int \mathrm{d}^{D+1}x \sqrt{(-1)^D\hat g}\, \left[ -\frac{1}{2}\hat m^2\hat \Phi^2+\frac{\hat\mu}{3!}\hat\Phi^3-\frac{\hat\lambda}{4!}\hat\Phi^4 \right]. 
	\end{equation}
	Here, $\hat m$  has mass dimension one, $\hat\mu$ has dimension $3-\frac{D+1}{2}$ and $\lambda$ has dimension $4-(D+1)$.
	Using the ansatz \eqref{scalar decomposition}, it is straightforward to see that the action takes the form 
	\begin{align}
		\mathcal S&=\mathcal S_f+\mathcal S_{int} \nonumber \\
		&=\int \mathrm{d}^Dx\sqrt{(-1)^{D-1}g}\left\{ \frac{R}{2\kappa^2}+ \frac{1}{2}(\partial \phi)^2-\frac{1}{4}e^{-2\sqrt\frac{D-1}{D-2}\kappa{\phi}}F^2+\frac{1}{2}\partial_\mu\varphi_0\partial^\mu\varphi_0 -\frac{1}{2}e^{\frac{2}{\sqrt{(D-1)(D-2)}}\kappa{\phi}}\hat m^2\varphi_0^2 \right.
		\nonumber \\
		& \hspace{3.8cm} \left. +\sum_{n=1}^{\infty} \partial_\mu\varphi_n\partial^\mu\varphi_n^*  -\sum_{n=1}^{\infty}\left(e^{2\sqrt{\frac{D-1}{D-2}}\kappa{\phi}} \frac{n^2}{L^2} +e^{\frac{2}{\sqrt{(D-1)(D-2)}}\kappa {\phi}}\hat m^2\right)\varphi_n\varphi_n^* 
		\right. 
		\nonumber \\
		&\hspace{3.8cm} +\sum_{n=1}^\infty \left[i{\sqrt{2}}\kappa \frac{n}{L}A^{\mu}\left(\partial_{\mu}\varphi_n \varphi_n^*-\varphi_n\partial_\mu\varphi_n^*\right)+{2}\kappa^2 \frac{n^2}{L^2}A_\mu A^\mu\varphi_n\varphi_n^* \right]
		\nonumber \\
		&\hspace{3.8cm} +e^{\frac{2}{\sqrt{(D-1)(D-2)}}\kappa {\phi}}
		\Bigg[ \frac{\mu}{3!}\varphi_0^3-\frac{\lambda}{4!}\varphi_0^4 +\mu\varphi_0\sum_{n=1}^\infty\varphi_n\varphi_n^*-\frac{\lambda}{2}\varphi_0^2\sum_{n=1}^\infty\varphi_n\varphi_n^*
		\nonumber \\
		&\hspace{3cm} -\frac{\lambda}{2}\varphi_0\sum_{m,n=1}^{\infty}\left(\varphi_m\varphi_m\varphi^*_{n+m} +\varphi_m^*\varphi_n^*\varphi_{n+m}\right) +\frac{\mu}{2}\sum_{n,m=1}^\infty \left( \varphi_n\varphi_m\varphi_{n+m}^*+\varphi_n^*\varphi_m^*\varphi_{n+m}\right) 
		\nonumber \\
		& \hspace{3cm} -\frac{\lambda}{3!}\sum_{m,n,p=1}^\infty\left(\varphi_m\varphi_n\varphi_p\varphi^*_{m+n+p}+\varphi^*_m\varphi^*_n\varphi^*_p\varphi_{m+n+p}\right)
		\nonumber\\
		& \hspace{4cm}-\frac{\lambda}{2}\sum_{n=1}^{\infty}\varphi_n\varphi_n^*\sum_{m=1}^{\infty}\varphi_m\varphi_m^*-\frac{\lambda}{4}\sum_{\underset{m\ne p,n\ne p ; m+n>p}{m,n,p=1}}^\infty\varphi_m\varphi_n\varphi_p^*\varphi_{n+m-p}^*\Bigg]\Bigg\},
	\end{align}
	where we have kept the notation compact, but, in our perturbative analysis, the dilaton will again be expanded around a background value $\phi_0$ as above. The couplings constants $\mu$ and $\lambda$ are defined, from their higher dimensional counterpart, as 
	\begin{equation}
		\mu=\frac{\hat\mu}{\sqrt{2\pi L}}, \qquad \lambda=\frac{\hat \lambda}{2\pi L}. 
	\end{equation}
	The tree-level masses for the zero mode $\varphi_0$ and the KK excitations are given by:
	\begin{equation}
		m^2_0= e^{\frac{2}{\sqrt{(D-1)(D-2)}}\kappa{\phi_0}}\hat m^2, \qquad  m_n^2= e^{2\sqrt{\frac{D-1}{D-2}}\kappa {\phi_0}}\frac{n^2}{L^2} +e^{\frac{2}{\sqrt{(D-1)(D-2)}}\kappa{\phi_0}}\hat m^2 . 
	\end{equation}
	%

	%%%%%%%%%%%%%%%%%%%%%%%%%%%%%%%%%%%%%%%%%%%%%%%%%%%%%%%%%%%%%%%%%%%%%%%%%%%%%%%%%%%%%

	%%%%%%%%%%%%%%%%%%%%%%%%%%%%%%%%%%%%%%%%%%%%%%%%%%%%%%%%%%%%%%%%%%%%%%%%%%%%%%%%%%%%%

	%%%%%%%%%%%%%%%%%%%%%%%%%%%%%%%%%%%%%%%%%%%%%%%%%%%%%%%%%%%%%%%%%%%%%%%%%%%%%%%%%%%%%

	\subsection{The Scalar Weak Gravity Conjecture} 
	\label{scatteringamplitudessection}
	
	\begin{figure}
		\centering
		\includegraphics[scale=0.78]{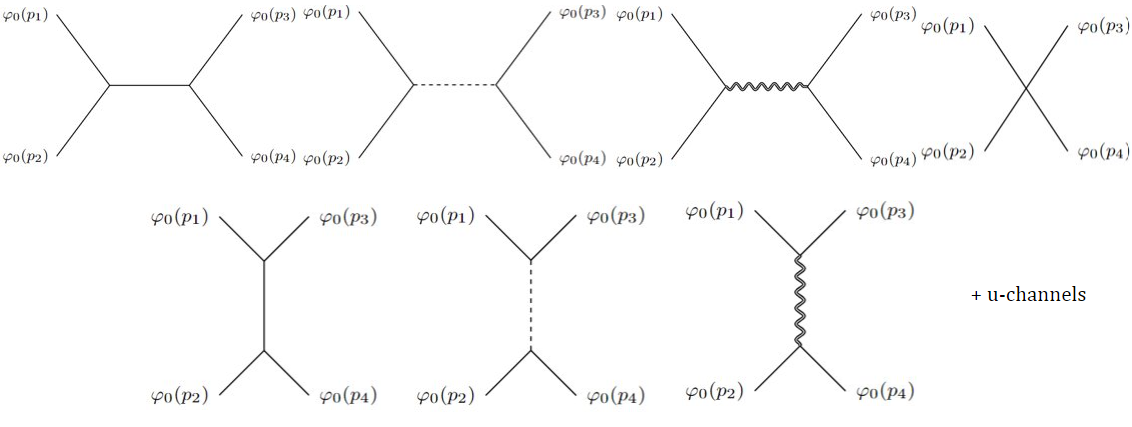}
		\caption{Feynman diagrams for the $\varphi_0\varphi_0\to\varphi_0\varphi_0$  scattering when a potential for the higher dimensional scalar, ``parent" of $\varphi_0$, has been turned o.n}
		\label{K_0 K_0}
	\end{figure}

	We start by computing the $\varphi_0\varphi_0\to\varphi_0\varphi_0$ amplitude. The diagrams intervening in the scattering are presented in the figure \ref{K_0 K_0}. The non-relativistic limit of the tree-level amplitude reads
	\begin{align}
		\label{phi0phi0tophi0phi0}
		i\mathcal M&=ie^{\frac{2}{\sqrt{(D-1)(D-2)}}\kappa{\phi_0}}\left[e^{\frac{2}{\sqrt{(D-1)(D-2)}}\kappa{\phi_0}}\frac{5}{3}\frac{\mu^2}{m_0^2}-\lambda\right] \nonumber \\
		&-\frac{i}{(D-1)(D-2)}\kappa^2 m_0^2-4\frac{i}{(D-1)(D-2)}\kappa^2 m_0^4\left(\frac{1}{t}+\frac{1}{u}\right) \nonumber \\
		&+i\frac{D-1}{D-2}\kappa^2 m_0^2-4i\frac{D-3}{D-2}\kappa^2 m_0^4\left(\frac{1}{t}+\frac{1}{u}\right),
	\end{align}
	where the different lines correspond to the contributions from the self-interaction, dilaton and graviton exchanges, respectively.   
	
	Following \cite{Benakli:2020pkm}, we compare the contributions to the amplitude at the energy scale given by the (massive) external states at rest. In the non-relativistic limit, we can further split \eqref{phi0phi0tophi0phi0} into contributions from short and long range interactions. We can identify an effective contact interaction:
	\begin{align}
		\label{CTPhi_0}
		i\mathcal M_{CT}^{(D)}&= ie^{\frac{2}{\sqrt{(D-1)(D-2)}}\kappa\phi_0}\left(\frac{5}{3}\frac{\mu^2}{\hat m^2}-\lambda-\frac{1}{(D-1)(D-2)}\kappa^2 \hat m^2+\frac{D-1}{D-2}\kappa^2 \hat m^2\right) \nonumber \\
		&=i\frac{e^{\frac{2}{\sqrt{(D-1)(D-2)}}\kappa\phi_0}}{2\pi L}\left(\frac{5}{3}\frac{\hat \mu^2}{\hat m^2}-\hat\lambda+2\pi L\frac{D}{D-1}\kappa^2 \hat m^2\right).
	\end{align}
	where in the first line we can identify the contributions from the scalar interaction for the first two terms, then from the dilaton and graviton, respectively. Using \eqref{relation M_P different dimensions} and the $(D+1)$-gravitational coupling $\hat \kappa=\sqrt{2\pi L}\,\kappa$, the last term is recognized to be the gravitational s-channel contribution to the $\hat\Phi\hat\Phi\to\hat\Phi\hat\Phi$ scattering in $D+1$ dimensions: 
	\begin{equation}
		\label{CTPhi_0-1}
		i\mathcal M_{CT}^{(D+1)}= i\frac{e^{\frac{2}{\sqrt{(D-1)(D-2)}}\kappa\phi_0}}{2\pi L}\left(\frac{5}{3}\frac{\hat \mu^2}{\hat m^2}-\hat\lambda+ \frac{(D+1)-1}{(D+1)-2}\hat \kappa^2 \hat m^2\right).  
	\end{equation}
	The above equation illustrates the fact that constraining the scalar interactions of the field $\hat\Phi$ to be dominant with respect to gravity in $D+1$ dimensions is enough to ensure that the scalar interactions of the zero mode $\varphi_0$ are dominant with respect to the combination of gravitational and dilatonic contributions in $D$ dimensions. 
	In other words, the effective (tree-level) non-relativistic four-point function of the zero mode $\varphi_0$ that emerges in the reduced-dimensional theory is the same as the effective non-relativistic four-point coupling for the "parent" field $\hat \Phi$ in the higher-dimensional theory. Requiring that in such a contact term, the contributions of the $\hat \Phi$ self-interactions are the dominant ones in the $D+1$ dimensions automatically ensures that the same property holds for the $\varphi_0$ self-interactions with respect to the set of interactions that appear in the $D$ dimensional theory.

	It is interesting to observe that the higher dimensional result is recovered here thanks to a cancellation, rather than an addition, between the graviton and dilaton mediated diagrams. This is dictated by the form of the $D$-dependent coefficient $\gamma_s(D) \equiv(D-1)/(D-2)$ appearing in front of the graviton-mediated amplitude in the $s$-channel which decreases with $D$: $\gamma_s(D+1)<\gamma_s(D)$. The dimension-dependent factor appearing in the $t$ and $u$-channels, $\gamma_{t,u}(D) \equiv(D-3)/(D-2)$ vary in the opposite direction. In other words, the peculiar feature is that, for the contact terms, the spin-2 and spin-0 bosonic mediators give opposite contributions. This feature will also appear in the amplitudes computed with the non minimal coupling to gravity. As a consequence of particular interest in the case of a massive dilaton the higher dimensional sub-dominance of gravity does not imply that gravity by itself (i.e. without the dilaton) is subdominant in the lower dimensional theory too. This violation happens in the parametric region
	\begin{equation}
		\frac{D}{D-1}\hat \kappa^2\hat m^2\le \left|\frac{5}{3}\frac{\hat\mu^2}{\hat m^2}-\hat \lambda\right| \le \frac{D-1}{D-2} \hat \kappa^2\hat m^2,  
	\end{equation}
	which is an interval of lenght $\hat \kappa^2\hat m^2/(D-1)(D-2)$ inversely proportional to the dimension $D$.

	\begin{figure}
		\centering
		\includegraphics[scale=0.78]{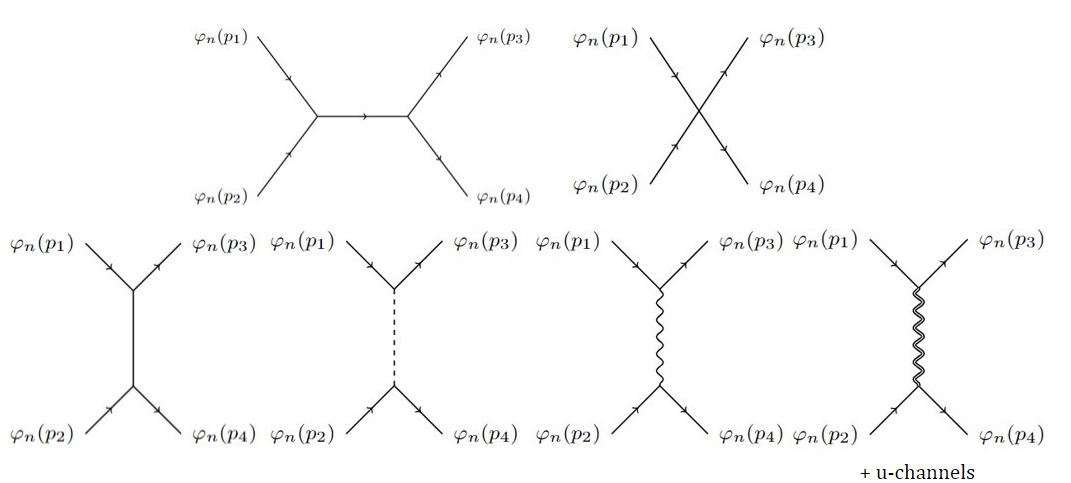}
		\caption{Feynman diagrams for the $\varphi_n\varphi_n\to\varphi_n\varphi_n$ scattering in the $t$-channel.}
		\label{KK}
	\end{figure}

	The amplitude $\varphi_n\varphi_n\to\varphi_n\varphi_n$ provides a generalization in the presence of self-interacting terms of the computation done in section \ref{SectionforcebetweenKKstates}. 
	The scattering amplitude receives contributions from gauge bosons, dilatons, gravitons in the t and u-channels, $\varphi_0$ exchange, from the s-channel exchange of a $\varphi_{2n}$ particle and from a 4-point contact term. These are the diagrams that are presented in figure \ref{KK} and lead to
	\begin{align}
		\label{phinphinscattering}
		i \mathcal M&= -ie^{\frac{4}{\sqrt{(D-1)(D-2)}}\kappa\phi_0}\mu^2\left(\frac{1}{s-m_{2n}^2}+\frac{1}{t-m_0^2}+\frac{1}{u-m_0^2}\right)-i\lambda e^{\frac{2}{\sqrt{(D-1)(D-2)}}\kappa\phi_0}   \nonumber \\
		&+i\left(\frac{1}{t}+\frac{1}{u}\right)\left(4g^2q_n^2m_n^2-4\frac{D-3}{D-2}\frac{m_n^4}{M_P^{D-2}}-(\partial_\phi m_n^2)^2\right)
	\end{align}
	with 
	\begin{equation}
		\partial_\phi m_n^2=\frac{1}{M_P^{(D-2)/2}}\left(\frac{2}{\sqrt{(D-1)(D-2)}}e^{\frac{2}{\sqrt{(D-1)(D-2)}}\kappa{\phi_0}}\hat m^2+2\sqrt{\frac{D-1}{D-2}}e^{2\sqrt{\frac{D-1}{D-2}}\kappa{\phi_0}}\frac{n^2}{L^2}\right).     
	\end{equation}

	%%%%%%%%%%%%%%%%%%%%%%%%%%%%%%%%%%%%%%%%%%%%%%%%%%%%%%%%%%%%%%%%%%%%%%%%%%%%%%%%%%%%%

	%%%%%%%%%%%%%%%%%%%%%%%%%%%%%%%%%%%%%%%%%%%%%%%%%%%%%%%%%%%%%%%%%%%%%%%%%%%%%%%%%%%%%

	\subsection{Massive dilatons}
	
	Let us consider for our illustrative discussion a simple potential for the dilaton in a polynomial expansion of the form
	\begin{equation}
		V(\phi)=\frac{1}{2}m^2_\phi \phi^2-\frac{\mu_\phi}{3!}\phi^3+\frac{\lambda_\phi}{4!} \phi^4.   
	\end{equation}
	
	In the $\varphi_0\varphi_0\to\varphi_0\varphi_0$ scattering amplitude \eqref{phi0phi0tophi0phi0}, the addition of a dilaton mass gives in the non-relativistic limit
	
	\begin{align}
		\label{phi_0 phi_0 with massive dilaton}
		i\mathcal M(\varphi_0\varphi_0\to\varphi_0\varphi_0)&=ie^{\frac{2}{\sqrt{(D-1)(D-2)}}\kappa{\phi_0}}\left[e^{\frac{2}{\sqrt{(D-1)(D-2)}}\kappa{\phi_0}}\frac{5}{3}\frac{\mu^2}{m_0^2}-\lambda\right] \nonumber \\
		&-4\frac{i}{(D-1)(D-2)}\kappa^2 m_0^4\frac{1}{s-m^2_\phi}-4\frac{i}{(D-1)(D-2)}\kappa^2 m_0^4\left(\frac{1}{t-m^2_\phi}+\frac{1}{u-m^2_\phi}\right) \nonumber \\
		&+i\frac{D-1}{D-2}\kappa^2 m_0^2-4i\frac{D-3}{D-2}\kappa^2 m_0^4\left(\frac{1}{t}+\frac{1}{u}\right),  
	\end{align}
	where the limit still needs to be implemented in the dilaton propagators according to its mass. We can thus follow the evolution of $\mathcal M$ with respect to $m_\phi$ to better expand it. \\

	For the $\varphi_n\varphi_n\to\varphi_n\varphi_n$ case, the scattering amplitude with the massive dilaton reads
	
	\begin{align}
		\label{amplitude phi_nphi_n dilaton massif}
		i\mathcal M(\varphi_n\varphi_n\to\varphi_n\varphi_n)&= -ie^{\frac{4}{\sqrt{(D-1)(D-2)}}\kappa{\phi_0}}\mu^2\left(\frac{1}{s-m_{2n}^2}+\frac{1}{t-m_0^2}+\frac{1}{u-m_0^2}\right)-i\lambda e^{\frac{2}{\sqrt{(D-1)(D-2)}}\kappa{\phi_0}}   \nonumber \\
		&-i(\partial_\phi m_n^2)^2\left(\frac{1}{t-m^2_\phi}+\frac{1}{u-m^2_\phi}\right)+i\left(\frac{1}{t}+\frac{1}{u}\right)\left(4g^2q_n^2m_n^2-4\frac{D-3}{D-2}\kappa^2 m_n^4\right).
	\end{align}

	%For later convenience, we define here $\bar m_n^2\equiv e^{2\sqrt{\frac{D-1}{D-2}}\kappa\phi_0}{n^2}/{L^2}$, so that $m_n^2=m_0^2+\bar m_n^2$.\\
	
	Putting all the analysis for both the $\varphi_0\varphi_0\to\varphi_0\varphi_0$ and  $\varphi_n\varphi_n\to\varphi_n\varphi_n$ scattering amplitudes together, we give a brief overview of the results here.\\
	
	When the mass $m_\phi$ of the dilaton is less than that of the zero mode, $m_0$, its mass can be neglected to first order in an expansion, in powers of $m_\phi$ over the exchanged momentum, and requiring that the self-interactions of a scalar field dominate in $D+1$ dimensions is sufficient to ensure that the same property is verified by its zero mode in $D$ dimensions; a result that follows from the studies of the previous sections. As soon as the mass of the dilaton is comparable to that of the $0$-mode, the massless dilaton approximation is no longer adequate and an appropriate discussion must be made for different denominators involving $m_\phi$, $m_0$, $m_n$ and $m_{2n}$. The analysis can be done easily but it is cumbersome and not really illuminating. In short, there is no easy way to relate combinations appearing in $D$ dimensions in this case with quantities already constrained, by assumption, in $D+1$ dimensions.

	%%%%%%%%%%%%%%%%%%%%%%%%%%%%%%%%%%%%%%%%%%%%%%%%%%%%%%%%%%%%%%%%%%%%%%%%%%%%%%%%%%%%%

	%%%%%%%%%%%%%%%%%%%%%%%%%%%%%%%%%%%%%%%%%%%%%%%%%%%%%%%%%%%%%%%%%%%%%%%%%%%%%%%%%%%%%
	
	%%%%%%%%%%%%%%%%%%%%%%%%%%%%%%%%%%%%%%%%%%%%%%%%%%%%%%%%%%%%%%%%%%%%%%%%%%%%%%%%%%%%%

	%%%%%%%%%%%%%%%%%%%%%%%%%%%%%%%%%%%%%%%%%%%%%%%%%%%%%%%%%%%%%%%%%%%%%%%%%%%%%%%%%%%%%
	
	\section{$\hat\Phi^2 R$ interaction}
	\label{section phi^2 R interaction}
	
	Let us consider now the effect on the different D-dimensional amplitudes of the presence of a non-minimal coupling to gravity of the form
	
	\begin{equation}
		\label{non-minimal}
		S_{(\xi)}=\int d^{D+1}x \sqrt{(-1)^D\hat g}\, \frac{\xi}{2} \hat\Phi^2 \hat R,
	\end{equation}
	with $\hat R$ the Ricci scalar (see for example \cite{Callan:1970ze}). We assume here that $\braket{\hat\Phi} = 0$ as a non-vanishing vev would correspond to a redefinition of the Planck mass and a shift of the canonical fields.
	After compactification, one gets:
	\begin{align}
		\label{phi2 R after compactification}
		S_{(\xi)}=\int d^{D}x \sqrt{(-1)^{D-1} g}\,    & \left[
		\xi\left(R-\kappa^2(\partial\phi)^2-\frac{2\kappa}{\sqrt{(D-1)(D-2)}}\nabla_\mu\partial^\mu\phi-\frac{1}{2}e^{-2\sqrt{\frac{D-1}{D-2}}\kappa\phi}\kappa^2F^2\right) \right. \nonumber \\  & \left. \times \left(\frac{\varphi_0^2}{2}+\sum_{n=1}^{\infty}\varphi_n\varphi_n^*\right) \right] .
	\end{align}
	This leads to new three-point couplings. First, using the linear expansion of the metric $g_{\mu\nu}=\eta_{\mu\nu}+2\kappa h_{\mu\nu}$, the $R$ term gives the new coupling $\kappa(\partial_\mu\partial_\lambda h^{\mu\lambda}-\Box h^\lambda_{\,\lambda})\left(\varphi_0^2+2\sum_{n=1}^{\infty}\varphi_n\varphi_n^*\right)$ of the graviton to the scalar matter fields.
	Then, the  $\nabla_\mu\partial^\mu\phi$ term, that we discard in previous sections as it takes the form of a total derivative, gives an additional three-point vertex between the dilaton and the matter fields and can enter, for example, in the computation of the dilatonic force in the non-relativistic limit.
	At first order in $\kappa$, we can write $\kappa{\nabla_\mu\partial^\mu\phi}=\kappa{\partial_\mu\partial^\mu\phi}+\mathcal {\cal O}(\kappa^2)$, the Christoffel symbols starting themselves at order $\kappa$.
	
	The $\varphi_0\varphi_0\to\varphi_0\varphi_0$ amplitude resulting from the action \eqref{phi2 R after compactification},  receives a contribution from the dilaton exchange (see Appendix \ref{Appendix A} for some details on the Feynman rules for two-derivative vertices)
	\begin{equation}
		i\mathcal M_\phi=-i\frac{4}{(D-1)(D-2)}\xi^2\kappa^2(s+t+u)=-i \frac{16}{(D-1)(D-2)}\xi^2\kappa^2m_0^2.
	\end{equation}
	and one from the graviton
	\begin{equation}
		\label{gravitational amplitude two non minimal vertices}
		i\mathcal M_G=4i\frac{D-1}{D-2}\xi^2\kappa^2(s+t+u)= 16i\frac{D-1}{D-2}\xi^2\kappa^2m_0^2.
	\end{equation}
	Their sum gives 
	
	\begin{equation}
		i\mathcal M_{(non-minimal)}=i\mathcal M_\phi +i\mathcal M_G= 4i\frac{D}{D-1}\xi^2\kappa^2(s+t+u)= 16i\frac{D}{D-1}\xi^2\kappa^2m_0^2.
	\end{equation}
	This matches the result one would obtain for the $\hat \Phi\hat \Phi\to \hat \Phi\hat \Phi$ scattering in $D+1$ dimensions. 
	
	At this point, we have computed tree-level four point amplitudes where both vertices arise either from minimal  or non-minimal couplings to gravity in D+1 dimensions. In order to compute the total  $\varphi_0\varphi_0\to\varphi_0\varphi_0$ amplitude we need to compute the contribution from ``mixed" diagrams involving one minimal and one non-minimal vertices. This mixed gravitational diagrams give in the $s$-channel 
	\begin{align}
		i\mathcal M^{\mathrm{G-mix.}}_\mathrm{s-channel}=-2i\xi\frac{\kappa^2}{s}\Bigg(&2\,p_1\cdot(p_1+p_2)\,p_2\cdot (p_1+p_2)-(p_1+p_2)^2(p_1\cdot p_2+m_0^2)+\frac{2}{D-2}(p_1+p_2)^2(p_1\cdot p_2)\nonumber\\
		&-\frac{D}{D-2}(p_1+p_2)^2(p_1\cdot p_2+m_0^2)\Bigg) 
	\end{align}
	and in the $t$-channel 
	\begin{align}
		i\mathcal M^{\mathrm{G-mix.}}_\mathrm{t-channel}=2i\xi\frac{\kappa^2}{t}\Bigg(&2\, p_1\cdot(p_1-p_3)\,p_3\cdot (p_1-p_3)-(p_1-p_3)^2(p_1\cdot p_3-m_0^2)+\frac{2}{D-2}(p_1-p_3)^2(p_1\cdot p_3)\nonumber\\
		&-\frac{D}{D-2}(p_1-p_3)^2(p_1\cdot p_3-m_0^2)\Bigg),
	\end{align}
	while the $u$ channel can be obtained through the replacements $t \leftrightarrow u $ and $p_3 \leftrightarrow p_4$.
	%In $D=4$ dimensions in the center of mass frame the amplitude reads
	%\begin{align}
	%i\mathcal M^{\mathrm{G-mix.}}_s=i\kappa^2\xi&(s+2m_0^2); \quad i\mathcal M^{\mathrm{G-mix.}}_{t}=i\kappa^2\xi(t+2m_0^2); \quad  i\mathcal M^{\mathrm{G-mix.}}_{u}=i\kappa^2\xi(u+2m_0^2) \nonumber\\
	%&\Longrightarrow i\mathcal M^{\mathrm{G-mix.}}=-i\kappa^2\xi (s+t+u+6m_0^2)=-10i\kappa^2\xi m_0^2.
	%\end{align}
	After some simple algebra, their sum reads
	\begin{align}
		\label{gravitational amplitude one non minimal vertex}
		i\mathcal M^{\mathrm{G-mix.}}_\mathrm{s-channel}=i\xi\kappa^2&\left(s+\frac{4 m_0^2}{D-2}\right); \, i\mathcal M^{\mathrm{G-mix.}}_\mathrm{t-channel}=i\xi\kappa^2\left(t+\frac{4 m_0^2}{D-2} \right); \, i\mathcal M^{\mathrm{G-mix.}}_\mathrm{u-channel}=i\xi\kappa^2\left(u+\frac{4m_0^2}{D-2} \right) \nonumber\\
		&\Longrightarrow i\mathcal M^{\mathrm{G-mix.}}=i\xi\kappa^2 \left(s+t+u+\frac{12}{D-2}m_0^2\right)=4i\xi\kappa^2\frac{D+1}{D-2} m_0^2.
	\end{align}
	The computation of the similar mixed diagrams with dilaton exchange  gives
	
	\begin{equation}
		i\mathcal M^{\mathrm{\phi-mix.}}=-12i\xi\kappa^2\frac{m_0^2}{(D-1)(D-2)},
	\end{equation}
	where each channel contributes the same amount. 
	
	Summing up all the contributions, the final result for the amplitude is
	
	\begin{equation}
		i\mathcal M^{\mathrm{mix.}}= 4i\xi\kappa^2 \frac{D+2}{D-1}m_0^2,
	\end{equation}
	as it is expected from the higher dimensional Lagrangian. Again, the higher dimensional gravitational contribution is obtained after a cancellation between the effective spin-2 and spin-0 mediators. 
	From the two results obtained above, we see that the direct non-minimal coupling to gravity \eqref{non-minimal} contributes with a constant term in the $\varphi_0\varphi_0\to \varphi_0\varphi_0$ amplitude.   
	If one takes the non-minimal coupling into account from the start and modifies the SWGC in $D$ generic dimensions requiring
	
	\begin{equation}
		\left|\frac{5}{3}\frac{\hat\mu^2}{ \hat m^2}-\lambda\right|\ge \left(\frac{D-1}{D-2}+4\xi \frac{D+1}{D-2}+16\xi^2\frac{D-1}{D-2}\right)\hat\kappa^2 \hat m^2, 
	\end{equation}
	the same property will be respected by the zero mode $\varphi_0$ in $D-1$ dimensions with the replacement of hatted by unhatted quantities ${\hat\mu^2}, \cdots \rightarrow \mu^2, \cdots$. \\ 
	\begin{comment}
		Concerning long range forces, even in the case with $V(\hat \Phi)=0$, the addition of the non-minimal coupling trivially makes the overall force attractive for the KK states. For a generic particle of mass $m$ and charge $q$ in a $U(1)$ gauge theory in arbitrary $d$ dimensions the requirement for a repulsive force takes the form
		
		\begin{equation}
			\label{WGC with non-minimal coupling}
			\hat g^2q^2\ge \left(\frac{d-3}{d-2}+\frac{\xi}{d-2} \right)\hat\kappa^2\hat m^2
		\end{equation}
		and is stable under the dimensional reduction to $d-1$ dimensions. 
		Similarly, this same $\xi$-dependent term should then be considered in the dilatonic WGC \eqref{generic dilatonic WGC} and in the WGC with generic scalar fields. \\
	\end{comment}
	
	In the  $\varphi_0\varphi_0\to \varphi_0\varphi_0$ scattering, the four point amplitudes appear as a sum of the three channels  $s,t,u$ whose coefficients add-up to a factor $s+t+u= 4m_0^2$. Therefore, the total amplitude does not increase with the exchanged momentum. This is not always the case as for example in the two examples of the $\varphi_n\varphi_n\to \varphi_n\varphi_n$ or $\varphi_n\varphi_n^*\to \varphi_n\varphi_n^*$ scattering amplitudes. The computation of the available channels, $t$ and $u$ in the first case, $s$ and $t$ in the second, proceeds as in the $\varphi_0$ case described above, but these contributions with two or one non minimal vertex do not close the sum $s+t+u$, as was the case in \eqref{gravitational amplitude two non minimal vertices} and \eqref{gravitational amplitude one non minimal vertex}.

	\begin{comment}
		To conclude, let us mention that application of the bound \eqref{WGC with non-minimal coupling} to magnetic monopoles would suggest the modification of the magnetic WGC with a new upper bound on the theory's cutoff: 
		
		\begin{equation}
			\label{modified magnetic WGC}
			\Lambda \lesssim \sqrt{\frac{1+\xi}{2}}\, gM_P.
		\end{equation}
	\end{comment}
	%%%%%%%%%%%%%%%%%%%%%%%%%%%%%%%%%%%%%%%%%%%%%%%%%%%%%%%%%%%%%%%%%%%%%%%%%%%%%%%%%%%%%

	%%%%%%%%%%%%%%%%%%%%%%%%%%%%%%%%%%%%%%%%%%%%%%%%%%%%%%%%%%%%%%%%%%%%%%%%%%%%%%%%%%%%%

	%%%%%%%%%%%%%%%%%%%%%%%%%%%%%%%%%%%%%%%%%%%%%%%%%%%%%%%%%%%%%%%%%%%%%%%%%%%%%%%%%%%%%

	%%%%%%%%%%%%%%%%%%%%%%%%%%%%%%%%%%%%%%%%%%%%%%%%%%%%%%%%%%%%%%%%%%%%%%%%%%%%%%%%%%%%%

	%%%%%%%%%%%%%%%%%%%%%%%%%%%%%%%%%%%%%%%%%%%%%%%%%%%%%%%%%%%%%%%%%%%%%%%%%%%%%%%%%%%%%

	\section{Higher dimensional gauge theory}
	
	So far, we have considered gravitational and scalar interactions in the higher dimensional theory. We will discuss now the case with gauge interactions. We consider a charged scalar $\hat \Phi$ of charge $q$ and mass $\hat M$ minimally coupled to a $U(1)$ gauge field $\hat B_M$ with gauge coupling $\hat g$ in $D+1$ dimensions  
	\begin{equation}
		\label{complex scalar action}
		\mathcal{S}_{EH,\Phi,H}^{(D+1)}=\int\mathrm{d}^{D+1}x\,\,\sqrt{(-1)^D\hat{g}}\,\, \left\{ \frac{\hat R}{2\hat\kappa^2}+\hat D_M \hat\Phi \hat D^M \hat\Phi^*-\hat M^2\hat\Phi\hat\Phi^* -\frac{1}{4}\hat H_{M\,N}\hat H^{M\,N}\right\},
	\end{equation}
	where $\hat H$ is the field strenght for the gauge field $\hat B$ and $\hat D_M$ the $D+1$ dimensional covariant derivative $\hat D_M\equiv \partial_M-i\hat g' q \hat B_M$, with $\hat g'$ the gauge coupling. For simplicity, we choose the following periodicities for the fields
	\begin{align} 
		\hat B_M(x,z+2\pi L)&=\hat B_M(x,z), \qquad \qquad \hat B_M(x,z)=\frac{1}{\sqrt{2\pi L}}\sum_{n=-\infty}^{+\infty}B_{(n) M}(x) e^{\frac{inz}{L}} \nonumber \\
		\hat \Phi(x,z+2\pi L)&=e^{i2\pi q_\Phi}\hat \Phi(x,z), \qquad \hat \Phi (x,z)=\frac{1}{\sqrt{2\pi L}}\sum_{n=-\infty}^{+\infty} \varphi_n(x)e^{i(n+q_\Phi)\frac{z}{L}},
	\end{align}
	where $q_\Phi$ is a putative charge of $\hat \Phi$ under an internal symmetry.
	The compactification of the (kinetic term of the) gauge field gives the lagrangian 
	\begin{align}
		\label{compactification gauge field}
		\mathcal L^{(D)}_H=&-e^{-2\alpha\phi}\left(\frac{H_0^{\,2}}{4}+\sum_{n=1}^\infty\frac{ |H_{(n)}|^{\,2}}{2}\right)+e^{-2\beta\phi}\left(\frac{(\partial h_0)^2}{2}+\sum_{n=1}^\infty\left|\partial h_n-i\frac{n}{L}B_{(n)}\right|^2\right)  \nonumber \\ &  +e^{-2\alpha\phi} A^\mu\left(-H_{(0) \mu\nu}\,\partial^\nu h_0+\sum_{n=1}^{\infty}H_{(n) \mu\nu}\left(\partial^\nu h_n^*-i\frac{n}{L}B_{(n)}^{*\, \nu}\right)+H^*_{(n) \mu\nu}\left(\partial^\nu h_n-i\frac{n}{L}B_{(n)}^{\, \nu}\right)\right)  \nonumber \\
		& +e^{-2\alpha\phi}\Bigg[A^2\left(\frac{\left(\partial h_0\right)^2}{2}+\sum_{n=1}^{\infty}\left|\partial h_n-i\frac{n}{L}B_{(n)}\right|^2\right)   \\
		&\qquad \qquad+A^\mu A^\nu\left(\partial_\mu h_0\partial_\nu h_0+2\sum_{n=1}^\infty\left(\partial_\mu h_n-i\frac{n}{L}B_{(n) \mu}\right)\left(\partial_\nu h_{n}-i\frac{n}{L}B_{(n) \nu}\right)^*\right)\Bigg], \nonumber 
	\end{align}
	where $h_0\equiv B_{(0) z}$ is a real scalar corresponding to the zero mode of the gauge field $\hat B_M$ component along the compact dimension $z$ and $h_n\equiv B_{(n) z}$ are the complex scalars forming the KK tower of the same field. From the above action, each field $h_n$ is seen to generate a mass for the KK excitations $B_{(n) \mu}$ of the non-compact components of the gauge field, that are then complex massive vectors, and to behave as the Goldstones in the Higgs mechanism (or in a Stuckelberg mechanism). Note that the relations 
	$B_{(-n) \mu}=B_{(n) \mu}^*$ and $h_{-n}=h_n^*$ are valid, although the same cannot be said for the Fourier modes of the complex field $\hat\Phi$.\\
	The $D$-dimensional lagrangian obtained from the kinetic and mass term of the scalar field $\hat \Phi$ reads 
	\begin{align}
		\label{action D dimensions for compactified complex scalar}
		\mathcal L^{(D)}_\Phi=&\sum_{n=-\infty}^{+\infty}\left|D\varphi_n\right|^2-\left(e^{2\alpha\phi}\hat M^2+e^{-2(\beta-\alpha)\phi}\left[\frac{n+q_\Phi}{L}-g'qh_0\right]^2\right)\left|\varphi_n\right|^2 \nonumber \\& +g'q\sum_{\underset{n\ne 0}{n,p=-\infty}}^{+\infty}\left[iB^\mu_{(n)}\left(\varphi_p\partial_\mu\varphi_{n+p}^*-\partial_\mu \varphi_p\varphi^*_{n+p}\right)-2g'qB_{(0)\mu}B_{(n)}^\mu \varphi_p\varphi_{n+p}^* \right.  \nonumber\\ 
		&\left. -g'q\sum_{\underset{m\ne 0}{m=-\infty}}^{+\infty}B_{(n) \mu}B_{(m)}^\mu\varphi_p\varphi^*_{n+m+p}\right]  \nonumber \\
		& +g'q A^\mu\left(\sum_{\underset{n\ne 0}{n,p=-\infty}}^{+\infty}i h_n\left(\partial_\mu \varphi_p\,\varphi^*_{n+p}-\varphi_p\partial_\mu\varphi^*_{n+p}\right)-2\sum_{n,p=-\infty}^{+\infty}\frac{n+p+q_\varphi}{L}\,B_{(n) \mu} \varphi_p\varphi^*_{n+p} \right. \nonumber \\
		&\left.\qquad\qquad+2 g'q h_0\sum_{n,p=-\infty}^{+\infty}B_{(n) \mu}\varphi_p\varphi^*_{n+p}+2 g'q \sum_{\underset{m\ne 0}{n,m,p=-\infty}}^{+\infty}h_m B_{(n) \mu}\varphi_p\varphi^*_{n+p}\right)\nonumber\\
		&+\left(A^2+e^{-2(\beta-\alpha)\phi}\right)\left(2g'q\sum_{\underset{n\ne 0}{n,p=-\infty}}^{+\infty}\left[\frac{n+p+q_\varphi}{L}-g'qh_0\right]h_n\varphi_p\varphi^*_{n+p} \right. \nonumber\\
		&\left.\qquad\qquad\qquad\qquad\qquad-g'^2q^2\sum_{\underset{m\ne 0}{n,m,p=-\infty}}^{+\infty}h_n h_m \varphi_p\varphi^*_{n+m+p}\right),
	\end{align}
	where $g'q\equiv \hat g' q/\sqrt{2\pi L}$ and when acting on $\varphi_n$
	\begin{equation}
		D_\mu\equiv \partial_\mu-ig'qB_{(0) \mu}-ig\left[\left(\frac{n+q_\Phi}{L}-g'qh_0\right)\right]A_\mu,   
	\end{equation}
	from which one can read the charge under the graviphoton.
	The $h_0$ term in this expression is a manifestation of the Aharonov-Bohm effect for the Wilson line of $B_z$, $\oint_z B_z$.
	
	Here we are interested in comparing the different gravitational and non-gravitational long range classical interactions, which can be obtained from the $t$-channel amplitudes. The $t$-channel contribution to the $\varphi_n(p_1)\varphi_n(p_2)\to\varphi_n(p_3)\varphi_n(p_4)$ scattering amplitude is
	\begin{align}
		\label{scattering amplitude D dimensions charged scalar}
		i\mathcal M_n=&\frac{i}{t}\left(g'^2q^2e^{\frac{2}{\sqrt{(D-1)(D-2)}}\kappa\phi_0}+ 2 \kappa^2\left(\frac{n+q_\Phi}{L}-g' qe^{-\sqrt{\frac{D-2}{D-1}}\kappa\phi_0}\bar h_0\right)^2e^{2\sqrt\frac{D-1}{D-2}\kappa\phi_0}\right) \left(p_1+p_3\right)\cdot\left(p_2+p_4\right)  \nonumber \\
		&-\frac{i}{t}\left[4g'^2q^2\left(g'q\bar h_0 e^{\frac{2}{\sqrt{(D-1)(D-2)}}\kappa\phi_0}-\frac{n+q_\Phi}{L}e^{\frac{D}{\sqrt{(D-1)(D-2)}}\kappa\phi_0}\right)^2\right.\nonumber\\
		&\qquad\left.+\left(2\sqrt{\frac{D-1}{D-2}}\kappa\left(\frac{n+q_\Phi}{L}-g'q\bar h_0e^{-\sqrt\frac{D-2}{D-1}\kappa\phi_0}\right)^2e^{2\sqrt\frac{D-1}{D-2}\kappa\phi_0}\right.\right.\nonumber\\
		&\qquad\left.\left.+\frac{2}{\sqrt{(D-1)(D-2)}}\kappa \hat M^2 e^{\frac{2}{\sqrt{(D-1)(D-2)}}\kappa\phi_0}\right)^2\right],
	\end{align}
	where  we have omitted writing the gravitational contribution, to avoid lengthy expressions, only to reinsert it in the next step when we perform the non-relativistic limit.
	The mass of the $n$th KK state can be read from the first line of the action in \eqref{action D dimensions for compactified complex scalar}
	\begin{align}
		\label{KK mass D dimensions charged scalar}
		m_n^2=\left(\frac{n+q_\Phi}{L}-g' qe^{-\sqrt{\frac{D-2}{D-1}}\kappa\phi_0}\bar h_0\right)^2e^{2\sqrt\frac{D-1}{D-2}\kappa\phi_0} +{\hat M}^2 e^{\frac{2}{\sqrt{(D-1)(D-2)}}\kappa\phi_0}\
	\end{align}
	
	Let us first consider the simplest case where $q_\Phi=\bar h_0=\hat M=0$. In the non-relativistic limit, for $n\ne 0$, the coefficient of $\frac{1}{t}$ in the t-channel amplitude takes the form
	\begin{align}
		\mathcal M_n^{\mathrm{t-pole}}&=\left(g'^2q^2 e^{\frac{2}{\sqrt{(D-1)(D-2)}}\kappa\phi_0}+2 \kappa^2m_n^2\right)4m_n^2 -4g'^2q^2m_n^2e^{\frac{2}{\sqrt{(D-1)(D-2)}}\kappa\phi_0} \nonumber\\
		&\quad -4\frac{D-1}{D-2}\kappa^2m_n^4-4\frac{D-3}{D-2}\kappa^2m_n^4 \nonumber\\
		&=0,
	\end{align}
	where $m_n^2$ in this case is simply $m_n^2=e^{2\sqrt{\frac{D-1}{D-2}}\kappa\phi_0}n^2/L^2$ and the gravitational scattering has been reinserted. The vanishing amplitude results  from the (expected) two by two cancellation of  interactions for the massive KK modes: namely gravitational vs dilatonic and D-dimensional  gauge vs scalar from the (D+1)-direction gauge field component. The $n=0$ amplitude is different as the zero mode is massless with our specific choice. The non gravitational amplitude reads
	
	\begin{equation}
		i\mathcal M_0^{relativistic}=\frac{i}{t}g'^2q^2e^{\frac{2}{\sqrt{(D-1)(D-2)}}\kappa\phi_0}(p_1+p_3)\cdot(p_2+p_4).
	\end{equation}
	
	Let us now consider the case $q_\Phi\ne 0$. The zero mode is massive
	
	\begin{equation}
		m_0^2=e^{\sqrt{\frac{D-1}{D-2}}\kappa\phi_0}\frac{q_\Phi^2}{L^2},    
	\end{equation}
	and the corresponding four-point amplitude is given by (again, we do not include here the gravitational contribution whose expression for generic exchanger momenta is long and not very illuminating)
	
	\begin{align}
		\label{M zero mode amplitude}
		i\mathcal M_{0}=&\frac{i}{t}\left(g'^2q^2e^{\frac{2}{\sqrt{(D-1)(D-2)}}\kappa\phi_0}+ 2 \kappa^2\frac{q^2_\Phi}{L^2}e^{2\sqrt\frac{D-1}{D-2}\kappa\phi_0}\right) \left(p_1+p_3\right)\cdot\left(p_2+p_4\right)  \nonumber \\
		&-\frac{i}{t}\left[4g'^2q^2\frac{q_\Phi^2}{L^2}e^{2\frac{D}{\sqrt{(D-1)(D-2)}}\kappa\phi_0}+4\frac{D-1}{D-2}\kappa^2\frac{q_\Phi^2}{L^2}e^{2\sqrt\frac{D-1}{D-2}\kappa\phi_0}\right].
	\end{align}
	
	In the non-relativistic limit, the total amplitude obtained by adding the gravitational contribution to \eqref{M zero mode amplitude}, cancels. The non-periodicity, which makes the zero mode massive, also generates couplings at $h_0$ and $\phi$, whose exchanges cancel, respectively, the gauge and gravitational amplitudes of the zero mode. This is to be expected since integer values of $q_\Phi$ reshuffle the KK states; what was the zero mode becomes one of the massive modes for which we have seen that the total amplitude disappears.  It is immediate to verify that the same is true for generic $n\ne 0$, $\mathcal M_n$ remains null, and the same thing happens if one turns on $\bar h_0$, as can be easily verified. 
	
	We can now study the general case. It is immediately verified that, after some algebra, in the non-relativistic limit the scattering amplitude \eqref{scattering amplitude D dimensions charged scalar} simplifies to
	\begin{align}
		i\mathcal M_{NR}^{(D)}=&4i e^{\frac{4}{\sqrt{(D-1)(D-2)}}\kappa\phi_0}\hat M^2\left(g'^2q^2-\frac{D-2}{D-1}\kappa^2\hat M^2\right) \nonumber \\
		=&4i \frac{e^{\frac{4}{\sqrt{(D-1)(D-2)}}\kappa\phi_0}}{2\pi L}{\hat M^2}\left(\hat g'^2q^2-\frac{(D+1)-3}{(D+1)-2}\hat \kappa^2\hat M^2\right) \propto i\mathcal M_{NR}^{(D+1)}
	\end{align}
	where one recognizes in the combination inside the parenthesis the $D+1$ dimensional corresponding dependence. The $q_\Phi$ and $\bar h_0$ dependences cancel out to leave this simple expression only in terms of the higher dimensional mass and charge. 
	We conclude that the requirement that the state in $D+1$ dimensions feels a repulsive long range force ensures that the KK modes in $D$ dimensions also feel a repulsive long range force. 
	
	The mapping of the $D+1$ dimensional $U(1)$ WGC into the $D$ dimensional form of the conjecture with gauge and scalar fields was discussed in \cite{Heidenreich:2015nta} from the requirement of extremal black holes and black p-branes decays, leading to the establishment of the dilatonic WGC, and in \cite{Lust:2017wrl} for the special case of a five to four dimensional circle compactification retaining only the zero modes. The analysis presented here generalizes, from the standpoint of scattering amplitudes, the connection between these different forms of the conjecture to the case with several gauge and scalar fields with reasonings involving the whole Kaluza-Klein tower.

	%%%%%%%%%%%%%%%%%%%%%%%%%%%%%%%%%%%%%%%%%%%%%%%%%%%%%%%%%%%%%%%%%%%%%%%%%%%%%%%%%%%%%

	%%%%%%%%%%%%%%%%%%%%%%%%%%%%%%%%%%%%%%%%%%%%%%%%%%%%%%%%%%%%%%%%%%%%%%%%%%%%%%%%%%%%%

	\subsection{Effective potential for $h_0$}
	
	Finally, we comment on the confrontation of the effective one-loop potential for the Wilson line with the scalar WGC of \cite{Benakli:2020pkm}. The potential is generated by the integration of the KK excitations\footnote{We use here the results of the effective potentials investigated in details for example in \cite{Antoniadis:2001cv} and at the one-loop level in a type I non-supersymmetric string model in \cite{Antoniadis:2000tq}.}. In the case of a circle compactification from five to four dimensions, the potential takes the simple form
	\begin{equation}
		\label{effective potential wilson line}
		V_{\mathrm{eff}}(h_0)=-\frac{3}{64\pi^6L^4}\sum_{n=1}^\infty\frac{\cos\left(2\pi n g'qh_0 L\right)}{|n|^5}=-\frac{3 \left(\text{Li}_5\left(e^{-2 \pi i g' q h_0 L}\right)+\text{Li}_5\left(e^{2 \pi i g' q h_0 L }\right)\right)}{128 \pi ^6 L^4},
	\end{equation}
	where the symbols $\text{Li}_n$ denote the usual Polylogarithm functions defined as 
	\begin{equation}
		\text{Li}_n(x)=\sum_{k=1}^\infty\frac{x^k}{k^n}.
	\end{equation}
	
	For the Wilson line to satisfy the Scalar WGC inequality of \cite{Benakli:2020pkm} around a generic background value $\bar h_0$ (we indicate with $\eta$ the excitations around it, $h_0=\bar h_0+\eta$), one then needs
	\begin{equation}
		\label{inequality wilson line-1}
		L^2\ge \frac{3\kappa^2}{ 2\pi^2 g'^2q^2}\left[\frac{\text{Li}_3\left(e^{i x}\right)+\text{Li}_3\left(e^{-i x}\right)}{\left| \frac{20}{9}\frac{ \left(\text{Li}_2\left(e^{i x}\right)-\text{Li}_2\left(e^{-i x}\right)\right)^2}{\text{Li}_3\left(e^{i x}\right)+\text{Li}_3\left(e^{-i x}\right)}-\log\left(2-2 \cos x\right)\right|}\right].
	\end{equation}
	where $x$ is defined to be $x\equiv 2\pi g'q \bar h_0 L$, to be respected for $m^2_\eta>0$, while the inequality is trivially verified for $m^2_\eta<0$, but this case is of no interest.  
	In the inequality \eqref{inequality wilson line-1}, the factor inside the square parenthesis on the right hand side is periodic and reaches a maximal value around $0.6-0.7$ in the regions of parameters where $m^2_\eta> 0$. Taken to be approximately an order one, the gravitational sub-dominance is then realized around any background value $\bar h_0$ if \footnote{Note that $\frac{\kappa^2}{g'^2}=\frac{\hat \kappa^2}{\hat g'^2}$, so we can express the bound either in terms of five- or four-dimensional quantities in the same form.}
	
	\begin{equation}
		\label{inequality wilson line}
		%\label{lower bound radius}
		L^2\ge\frac{3\hat \kappa^2}{2\pi ^2 \hat g'^2q^2}=\frac{3}{2\pi ^2  g'^2q^2}\frac{1}{M_P^2},
	\end{equation}
	which means that the compactification length cannot be parametrically smaller than the Planck's one as expected. 
	
	From \eqref{compactification gauge field} and \eqref{action D dimensions for compactified complex scalar}, it is immediate to observe that the self-couplings induced by radiative corrections are not the only ones that can appear in the $4$-point function $\eta\eta\to\eta\eta$. A first contribution may come from the kinetic term of $h_0$, coupled to the dilaton as in \eqref{compactification gauge field}. This gives a two derivative vertex that would then induce contributions to the four point function proportional to the scalar product of external momenta ($p_1\cdot p_2\times p_3\cdot p_4$ in the s-channel, and so on). For the effective four point non relativistic coupling, this only accounts for a shift of the gravitational contribution, the second term in \eqref{inequality wilson line}. In particular,  the numerical coefficient $3/2$ should be changed with $5$ in \eqref{inequality wilson line} and all the subsequent inequalities.

	%%%%%%%%%%%%%%%%%%%%%%%%%%%%%%%%%%%

	%%%%%%%%%%%%%%%%%%%%%%%%%%%%%%%%%%%%%%%%%%%%%%%%%%%%%%%%%%%%%%%%%%%%%%%%%%%%%%%%%%%%%

	%%%%%%%%%%%%%%%%%%%%%%%%%%%%%%%%%%%%%%%%%%%%%%%%%%%%%%%%%%%%%%%%%%%%%%%%%%%%%%%%%%%%%

	\section{Conclusions}

	An extra dimension for our space-time was originally introduced to unify gravity with electromagnetism: \cite{Kaluza:1921tu,Klein:1926tv,Klein:1926fj,Einstein:1938fk}. From the point of view of a lower dimensional observer, this unification makes the KK modes undergo attractive gravitational plus scalar interactions and repulsive electric interactions with the same intensity. This motivated the use of the KK states interactions in this work to extract the form of the inequalities that appear when one is interested in comparing gravitational interactions to other types of interactions. 
	
	Taking into account the scalar interaction due to the presence of a dilaton, the calculation of four-point amplitudes allowed us to find the inequalities of the Dilatonic WGC. Our observations go further, with the extension of the construction to include interactions in the higher dimension, and we have shown how the Scalar WGC is found as well as the behavior of these conjectures under dimensional reduction. Meanwhile, we have also computed a number of scattering amplitudes for the pair production of KK states and have been able to compare the contributions of the different channels for spacetime dimensions $D \ge 4$.

	%%%%%%%%%%%%%%%%%%%%%%%%%%%%%% appendices%%%%%%%%%%%%%%%%%%%%%%%%%%%%%%%%%%%%%%%%%%%%%%%%%%
	
	\vskip 10 pt
	\appendix

	%%%%%%%%%%%%%%%%%%%%%%%%%%%%%%%%%%%%%%%%%%%%%%%%%%%%%%%%%%%%%%%%%%%%%%%%%%%%%%%%%%%%%

	%%%%%%%%%%%%%%%%%%%%%%%%%%%%%%%%%%%%%%%%%%%%%%%%%%%%%%%%%%%%%%%%%%%%%%%%%%%%%%%%%%%%%

	%%%%%%%%%%%%%%%%%%%%%%%%%%%%%%%%%%%%%%%%%%%%%%%%%%%%%%%%%%%%%%%%%%%%%%%%%%%%%%%%%%%%%
	
	\section{Lagrangians with derivative interactions}
	\label{Appendix A}

	One subtlety that we wish to address here is related to the nature and the use of derivative interactions in perturbation theory. The perturbative expansion is an expansion of the exponential $e^{-i\int d^Dx \mathcal H_I}$ in powers of $\mathcal H_I$, the interaction hamiltonian in the interaction picture. When the lagrangian presents derivative interactions, one should be careful to correctly construct $\mathcal H_I$ before announcing the Feynman rules. Interactions containing more than one derivative of fields can generate new genuine additional Feynman rules \cite{Gerstein:1971fm}. The analog of this result was found, in the path integral formalism, in \cite{Honerkamp:1996va}.  We illustrate this in two simple examples closely related to the cases studied. 
	
	\subsection{Interactions with derivatives of a gauge field}
	
	We first present the case of the theory defined by
	\begin{equation}
		\mathcal L=\frac{1}{2}\partial_\mu\phi\partial^\mu\phi-\frac{1}{4}e^{-2\sqrt{\frac{D-1}{D-2}}\kappa{\phi}}\left(\partial_\mu A_\nu-\partial_\nu A_\mu\right)\left(\partial^\mu A^\nu-\partial^\nu A^\mu\right).
	\end{equation}
	We have singled out here only the part of interest to us to highlight the interaction between the dilaton $\phi$ and derivatives of the graviphoton $A_\mu$.
	We will work in the usual radiation gauge $A_0=0, \vec\nabla\cdot \vec A=0$.
	Computation of the canonical conjugate momenta give us
	
	\begin{equation}
		\begin{cases}
			\Pi_{A_0}=0 \\
			\Pi_{A_i}=-\left(1+\sum_{m=1}^\infty\left(-2\sqrt{\frac{D-1}{D-2}}\kappa\right)^m\frac{\phi^m}{m!}\right)F^{0i} \\
			\Pi_\phi=\partial_0\phi.
		\end{cases}
	\end{equation}
	The fact that $\Pi_{A_0}=0$ is, of course, what we should expect in a canonical formalism. The Heisenberg picture hamiltonian is obtained as 
	\begin{align}
		\mathcal H &=\Pi_{A_\mu}\partial_0 A_\mu+\Pi_\phi\partial_0\phi-\mathcal L \nonumber \\
		&=-\frac{1}{2}F_{0i}F^{0i}+\frac{1}{4}F_{ij}F^{ij}+\frac{1}{2}\partial_0\phi\partial_0\phi+\frac{1}{2}\partial_i\phi\partial_i\phi-\sum_{m=1}^\infty \left(-2\sqrt{\frac{D-1}{D-2}}\kappa\right)^m\frac{\phi^m}{m!}\left(F^{0i}F_{0i}-\frac{F_{\mu\nu}F^{\mu\nu}}{4}\right) \nonumber \\
		&=\frac{1}{2}\Pi_{A_i}\Pi_{A_i}+\frac{1}{4}F_{ij}F^{ij}+\frac{1}{2}\Pi_\phi\Pi_\phi+\frac{1}{2}\partial_i\phi\partial_i\phi+\frac{1}{4}\sum_{m=1}^\infty \left(-2\sqrt{\frac{D-1}{D-2}}\kappa\right)^m\frac{\phi^m}{m!}F^{\mu\nu}F_{\mu\nu} \nonumber\\
		&\hspace{0.5 cm}+\frac{1}{2}\sum_{m=1}^\infty\left[ \left(-2\sqrt{\frac{D-1}{D-2}}\kappa\right)^m\frac{\phi^m}{m!}\right]^2 F^{0i}F_{0i}.
	\end{align}
	The transition to the interaction picture is done making the following replacements:
	\begin{equation}
		\begin{cases}
			\Pi_{A_i}\to -F^{0i}\,(=\Pi_{A_i,\,I})  \\
			F_{ij}\to F_{ij} \\
			F^{0i}\to F^{0i}\left(1+\sum_{m=1}^\infty \left(-2\sqrt{\frac{D-1}{D-2}}\kappa\right)^m\frac{\phi^m}{m!}\right)^{-1}\\
			\Pi_\phi\to\partial_0\phi\\
			\phi\to\phi \\
			\partial_0\phi\to\partial_0\phi
		\end{cases}
	\end{equation}
	Some simple algebra finally get us to the interaction picture hamiltonian in the form
	\begin{align}
		\mathcal H &=-\frac{1}{2}F_{0i}F^{0i}+\frac{1}{4}F_{ij}F^{ij}+\frac{1}{2}\partial_0\phi\partial_0\phi+\frac{1}{2}\partial_i\phi\partial_i\phi+\frac{1}{4}\sum_{m=1}^\infty \left(-2\sqrt{\frac{D-1}{D-2}}\kappa\right)^m\frac{\phi^m}{m!} F^{\mu\nu}F_{\mu\nu}\nonumber\\
		&\hspace{0.5 cm}-\frac{1}{2}\frac{\sum_{m=1}^\infty\left[ \left(-2\sqrt{\frac{D-1}{D-2}}\kappa\right)^m\frac{\phi^m}{m!}\right]^2}{1+\sum_{m=1}^\infty \left(-2\sqrt{\frac{D-1}{D-2}}\kappa\right)^m\frac{\phi^m}{m!}} F^{0i}F_{0i}.
	\end{align}
	Careful construction of the interaction hamiltonian reveals the presence of an additional term to the naive expectation, to the extent that 
	
	\begin{equation}
		\mathcal H_I=-\mathcal L_I-\frac{1}{2}\frac{\sum_{m=1}^\infty\left[ \left(-2\sqrt{\frac{D-1}{D-2}}\kappa\right)^m\frac{\phi^m}{m!}\right]^2}{1+\sum_{m=1}^\infty \left(-2\sqrt{\frac{D-1}{D-2}}\kappa\right)^m\frac{\phi^m}{m!}} F^{0i}F_{0i},
	\end{equation} 
	with the new term sharing the same structure with the one found in the model of \cite{Gerstein:1971fm}. \\
	Combining this result with the two derivative propagator\footnote{Given here in the covariant gauge, to keep a simple notation.}
	
	\begin{equation}
		\label{corrected gauge propagator}
		\braket{\partial_\mu A_\rho \partial_\nu A_\sigma}(q)=i\eta_{\rho\sigma}\frac{q_\mu q_\nu}{q^2(+i\epsilon)}-i\eta_{\rho\sigma}\eta_{\mu\,0}\eta_{\nu\,0}
	\end{equation}
	we finally have the explicit form of the non standard Feynman rules we should consider in the minimally coupled (i.e. with $\xi=0$) dimensionally reduced theory. The additional term consists in an infinite series in powers of $\kappa\phi$ starting at order $2$ and defining a vertex with two gauge bosons. As such, it will not enter any of the computations we have performed, but certainly need to be considered, alongside with the propagator corrections, even at tree level, when looking at different physical processes, like $\phi\phi\to \gamma\gamma$ and $\phi\gamma\to\phi\gamma$ ones. 
	
	\subsection{Toy model for the two-derivative interaction of the non-minimal coupling}
	\label{appendix two derivative interaction}
	
	The second model we present here aims to capture the main properties of the new vertices brought in by the non-minimal coupling to gravity. 
	We explicitly show, with the simplest toy model, that the different additional pieces due to such derivatives cancel each other, allowing the use of naive perturbation theory. \\
	
	Let us take, for definiteness, the following lagrangian:
	
	\begin{equation}
		\label{toy model lagrangian}
		\mathcal L=\frac{1}{2}(\partial\phi)^2 +\frac{1}{2}(\partial\varphi)^2+\frac{a}{2}\kappa (\partial^2\phi) \varphi^2+\frac{b}{2}\kappa^2  (\partial\phi)^2\varphi^2=\frac{1}{2}(\partial\phi)^2 +\frac{1}{2}(\partial\varphi)^2-a\kappa (\partial\phi\cdot \partial\varphi) \varphi+\frac{b}{2}\kappa^2  (\partial\phi)^2\varphi^2,
	\end{equation}
	where $a$ and $b$ are dimensionless constants. In keeping the parallel with the cases discussed in the text, one should think of $\phi$ as a massless mediator and $\varphi$ the matter field. The addition of a mass term for $\varphi$ does not change the computations.\\ 
	The conjugate momenta are 
	
	\begin{equation}
		\begin{cases}
			\Pi_\phi=\partial_0\phi -a\kappa\phi\partial_0\varphi +b\kappa^2\partial_0\phi\,\varphi^2 \\
			\Pi_\varphi=\partial_0\varphi -a\kappa\varphi\partial_0\phi,
		\end{cases}
	\end{equation}
	and, inverting the relations, we obtain
	
	\begin{equation}
		\begin{cases}
			\partial_0\phi =\frac{\Pi_\phi+ a\kappa\varphi\Pi_\varphi}{1+(b-a^2)\kappa^2\varphi^2} \\
			\partial_0\varphi=\Pi_\varphi+ a\kappa\varphi \frac{\Pi_\phi+a\kappa\varphi\Pi_\varphi}{1+(b-a^2)\kappa^2\varphi^2}. 
		\end{cases}
	\end{equation}
	Following the steps described above, the interaction picture hamiltonian is obtained:
	\begin{align}
		\label{full hamiltonian toy model}
		\mathcal H&=\frac{\partial_0\phi(\partial_0\phi+a\kappa\varphi\partial_0\varphi)}{1+(b-a^2)\kappa^2\varphi^2}+\partial_0\varphi\left(\partial_0\varphi+a\kappa\varphi\frac{\partial_0\phi+a\kappa\varphi\partial_0\varphi}{1+(b-a^2)\kappa^2\varphi^2}\right)-\frac 1 2 \left(\frac{\partial_0\phi+ a\kappa\varphi\partial_0\varphi}{1+(b-a^2)\kappa^2\varphi^2}\right)^2 \nonumber \\
		&-\frac 1 2 \left(\partial_0\varphi+a\kappa\varphi\frac{\partial_0\phi+a\kappa\varphi\partial_0\varphi}{1+(b-a^2)\kappa^2\varphi^2}\right)^2+\frac{1}{2}\partial_i\phi\partial_i\phi+\frac{1}{2}\partial_i\varphi\partial_i\varphi+a\kappa\varphi\partial_0\varphi\frac{\partial_0\phi+a\kappa\varphi\partial_0\varphi}{1+(b-a^2)\kappa^2\varphi^2} \nonumber \\
		&+a^2\kappa^2\varphi^2\left(\frac{\partial_0\phi+ a\kappa\varphi\partial_0\varphi}{1+(b-a^2)\kappa^2\varphi^2}\right)^2-\frac{b}{2}\kappa^2\varphi^2\left(\frac{\partial_0\phi+ a\kappa\varphi\partial_0\varphi}{1+(b-a^2)\kappa^2\varphi^2}\right)^2-a\kappa\varphi \partial_i\phi\partial_i\varphi+\frac b 2 \kappa^2\varphi^2\partial_i\phi\partial_i\phi 
	\end{align}
	
	Expanding to second order in $\kappa$, to match the usual contributions to the $\varphi\varphi\to \varphi\varphi$ or $\phi\phi\to \varphi\varphi$ amplitudes from \eqref{toy model lagrangian}, we get
	\begin{align}
		\label{second order hamiltonian toy model}
		\mathcal H&=\frac{1}{2}(\partial_0\phi\partial_0\phi+ \partial_i\phi\partial_i\phi)+\frac{1}{2}(\partial_0\varphi\partial_0\varphi+\partial_i\varphi\partial_i\varphi)+ a\kappa\varphi(\partial_0\varphi \partial_0 \phi-\partial_i\varphi \partial_i \phi)-\frac{b}{2} \kappa^2\varphi^2(\partial_0\phi\partial_0\phi- \partial_i\phi\partial_i\phi)\nonumber \\
		&+\frac{a^2}{2}\kappa^2\varphi^2\left(  \partial_0 \phi\partial_0\phi +\partial_0 \varphi\partial_0\varphi \right) +O\left(\kappa^3\right).    
	\end{align}
	We recognize, in the first line, the sum $\mathcal H_{\mathrm{free}}-\mathcal L_I$ that is usually found in perturbation theory with no derivative interactions. The operator in the second line, as well as all the higher orders ones that can be derived from \eqref{full hamiltonian toy model}, are due to the derivative interactions in \eqref{toy model lagrangian}. Equation \eqref{second order hamiltonian toy model} shows that, at the level of the interaction picture hamiltonian, we get additional 4-point vertices with respect to the usual ones. \\
	
	We now check the impact of such additional interactive terms through the explicit computation of the $\varphi(p_1)\varphi(p_2)\to \varphi(q_1)\varphi(q_2)$ scattering amplitude. Taking into account the corrections to the scalar propagator (analogous to \eqref{corrected gauge propagator}), the usual ($-\mathcal L_I$) interactions give, in each one of the $s,t$ and $u$ channels
	
	\begin{equation}
		i\mathcal M_{\mathrm{(-\mathcal L_I)}}=-ia^2\kappa^2P_\mu P_\nu\left(\frac{P^\mu P^\nu}{P^2}-\eta^\mu_0\eta^\nu_0\right),    
	\end{equation}
	where $P$ is the appropriate momentum factor in each channel ($P=p_1+p_2$, $P=p_1-p_3$ and $P=p_1-p_4$, respectively, in $s,t$ and $u$). After some algebra, the four $\varphi$ contact term in \eqref{second order hamiltonian toy model} accounts for a contribution
	\begin{equation}
		i\mathcal M_{\mathrm{contact}}=-2ia^2\kappa^2(p_{1,0}^2+p_{2,0}^2+q_{1,0}^2-q_{1,0}p_{1,0}),     
	\end{equation}
	where the notation $p_{i,0}$ means the zero component of the momentum $p_i$.
	
	Putting it all together one gets
	\begin{equation}
		i\mathcal M=-ia^2\kappa^2\Big\{s+t+u-\left((p_{1,0}+p_{2,0})^2+(p_{1,0}-q_{1,0})^2+(p_{1,0}-q_{2,0})^2\right)+2(p_{1,0}^2+p_{2,0}^2+q_{1,0}^2-q_{1,0}p_{1,0})\Big\}.
	\end{equation}
	Using momentum conservation one can show that, again after some algebra, the non covariant pieces cancel leaving the same result one would have guessed using the naive Feynman rules from the lagrangian \eqref{toy model lagrangian} associating the appropriate momentum factor to each derivative:
	\begin{equation}
		i\mathcal M=-ia^2\kappa^2\left(s+t+u\right).
	\end{equation}
	
	The type of vertices being the same, this same cancellation happens in the ``pair production"-like amplitude $\phi\phi\to\varphi\varphi$. 
	
	This toy model explicitly shows the cancellation between different non covariant pieces arising in the computation of amplitudes with two derivative vertices and justifies, a posteriori, the use of naive perturbation theory we made in section \ref{section phi^2 R interaction}.

	%%%%%%%%%%%%%%%%%%%%%%%%%%%%%%%%%%%%%%%%%%%%%%%%%%%%%%%%%%%%%%%%%%%%%%%%%%%%%%%%%%%%%

	%%%%%%%%%%%%%%%%%%%%%%%%%%%%%%%%%%%%%%%%%%%%%%%%%%%%%%%%%%%%%%%%%%%%%%%%%%%%%%%%%%%%%

	%%%%%%%%%%%%%%%%%%%%%%%%%%%%%%%%%%%%%%%%%%%%%%%%%%%%%%%%%%%%%%%%%%%%%%%%%%%%%%%%%%%%%

	%%%%%%%%%%%%%%%%%%%%%%%%%%%%%%%%%%%%%%%%%%%%%%%%%%%%%%%%%%%%%%%%%%%%%%%%%%%%%%%%%%%%%

	%%%%%%%%%%%%%%%%%%%%%%%%%%%%%%%%%%%%%%%%%%%%%%%%%%%%%%%%%%%%%%%%%%%%%%%%%%%%%%%%%%%%%

	%%%%%%%%%%%%%%%%%%%%%%%%%%%%%%%%%%%%%%%%%%%%%%%%%%%%%%%%%%%%%%%%%%%%%%%%%%%%%%%%%%%%%

	\section{Helicity basis and Mandelstam variables} 
	\label{Helicity method appendix}
	
	In the computation of the pair production diagrams, we need to  deal with external states polarizations for massless helicity-1 and helicity-2 particles. This is of no concern when we compute the squared amplitude, as it is usually treated by means of the replacements $\sum_{\mathrm{pol}}\epsilon_{\mu}(p)\epsilon^*_{\nu}(p)\rightarrow -g_{\mu \nu}$ for photon amplitudes and $\sum_{\mathrm{pol}}\epsilon_{\mu\nu}(p)\epsilon^*_{\rho\sigma}(p)=\sum_{\mathrm{pol}}\epsilon_\mu(p)\epsilon_\nu(p)\epsilon^*_\rho(p)\epsilon^*_\sigma(p)\rightarrow \mathcal P_{\mu\nu\rho\sigma}$ for graviton ones. If, on the other hand, we want to consider the amplitude more directly and not its square, we need to choose a basis for the polarizations and the momentum, and perform the calculations within this basis. 
	
	For the case of the pair production, the in-going states relevant here are either photons or gravitons, while the outgoing ones are massive particles. We perform here the computations in the center of momentum frame. 
	
	Starting from the $D=4$ case, we write the momenta
	\begin{align}
		p_1=E_p(1,0,0,1),\quad p_2=E_p(1,0,0,-1),\quad p_3=(E_p,p\sin\theta,0,p\cos\theta), \quad  p_4=(E_p,-p\sin\theta,0,-p\cos\theta)
	\end{align}   
	and the polarizations
	\begin{align}
		\epsilon_1^{\pm} \equiv \epsilon(p_1)^{\pm}=\frac{1}{\sqrt{2}}(0,\mp 1,-i,0),\qquad\epsilon_2^{\pm} \equiv \epsilon(p_2)^{\pm}=\frac{1}{\sqrt{2}}(0,\pm 1,-i,0).
	\end{align}
	The scalar products appearing in the amplitudes can now be explicitly performed in this particular basis and the results can then be rewritten in terms of the Mandelstam variables using the following relations:
	\begin{align}
		p^2=\frac{s-4m^2}{4},\quad \sin^2\theta=\frac{(t-u)^2}{s(s-4m^2)},\quad\cos^2\theta=\frac{4tu-4m^2}{s(s-4m^2)} 
	\end{align}
	At this point, we need to separate the contributions coming from different helicities. For definiteness, we refer now to the amplitude in \eqref{amplitude photon photon pair production}, that we report here for the reader's convenience
	\begin{align*}
		i\mathcal M_{\gamma\gamma}=&ig^2q_n^2\, \,   \epsilon_\mu(p_1)\epsilon_\nu(p_2)\left(\frac{(2 p_3^\mu-p_1^\mu) (2 p_4^\nu-p_2^\nu)}{t-m_n^2}+\frac{(2 p_4^\mu-p_1^\mu)(2 p_3^\nu-p_2^\nu)}{u-m_n^2}+2\eta^{\mu\nu}\right) \\ \nonumber &-2i g^2q_n^2\frac{D-1}{D-2}\, \, \, \epsilon_\mu(p_1)\epsilon_\nu(p_2)\frac{p_1\cdot p_2 \eta^{\mu\nu}-p_1^\nu p_2^\mu}{s}.
	\end{align*}
	A great simplification comes when we deal more directly with the amplitudes components. We can in fact use the property\footnote{When using the usual shortcut $\sum_{\rm pol}\epsilon_\mu(p)\epsilon_\nu(p)=-g_{\mu\nu}$ this simplification cannot be used.} $\epsilon(p)\cdot p=0$. With our choice of basis, we also have $\epsilon(p_1)\cdot p_2=\epsilon(p_2)\cdot p_1=0$, so that, for the purposes of the calculation with the helicity method, we can use the following expression for the amplitude
	\begin{align}
		\label{photon amplitude helicity}
		\mathcal M_{\gamma\gamma}=&4g^2q_n^2\, \,  \left\{\frac{\epsilon(p_1)\cdot p_3 \, \epsilon(p_2)\cdot p_4}{t-m_n^2}+\frac{\epsilon(p_1)\cdot p_4 \, \epsilon(p_2)\cdot p_3}{u-m_n^2}+\frac{\epsilon(p_1)\cdot\epsilon(p_2)}{2}\left(1-\frac{D-1}{D-2} \frac{p_1\cdot p_2}{s}\right)\right\}.
	\end{align}
	We denote with $\mathcal M_{\pm \pm}$ the different contributions, with the $\pm$ referring to the helicities of the polarization. We have then 
	\begin{align}
		i\mathcal M_{++}=2i(gq_n)^2\left(\frac{m_n^2s}{(t-m_n^2)(u-m_n^2)}-\gamma_d\frac{3}{4}\right), \qquad i\mathcal M_{+-}=-2i(gq_n)^2\frac{(m_n^4-ut)}{(t-m_n^2)(u-m_n^2)},
	\end{align}
	where we have introduced a factor $\gamma_d$ in front of the term arising from the dilaton such that we retrieve the result for our KK theory when $\gamma_d=1$ and the usual result for a $U(1)$ gauge theory when $\gamma_d=0$. 
	To compute the total amplitude, we average over the in-going polarizations and obtain
	in the threshold limit
	\begin{align}
		|\mathcal M_{\gamma\gamma}|^2&=\frac{1}{4}\left(2|\mathcal M_{++}|^2+2|\mathcal M_{+-}|^2\right)\to 2\left(1-\gamma_d\frac
		34\right)^2(gq_n)^4.
	\end{align}
	When $\gamma_d=0$, the overall numerical factor is $2$, while for $\gamma_d=1$, it is $1/8$, matching
	the results obtained in Section \ref{pair production subsection} for $D=4$. It is immediate to realize that, in the threshold limit, only the $\epsilon(p_1)\cdot \epsilon(p_2)$ term contributes. 
	
	The same method outlined above can be used for any other number of dimensions $D$, where the gauge bosons have $D-2$ independent helicity states. For instance, in the $D=5$ case, the helicity basis can be taken as 
	
	\begin{align}
		\epsilon_1^1&=\frac{1}{\sqrt{2}}(0,-1,-i,0,0) \qquad \,\epsilon_2^1= \frac{1}{\sqrt{2}}(0,1,-i,0,0) \nonumber \\
		\epsilon_1^2&=\frac{1}{\sqrt{2}}(0,1,-i,0,0) \qquad \quad \epsilon_2^2=\frac{1}{\sqrt{2}}(0,-1,-i,0,0) \nonumber \\
		\epsilon_1^3&=(0,0,0,1,0)\qquad \qquad\, \quad \epsilon_2^3=(0,0,0,-1,0). 	
	\end{align}
	For any $D>4$, the polarization basis can be chosen such that, for both $p_1$ and $p_2$, the first two polarizations are the same as in $D=4$, while the other polarizations are $\epsilon^i_1=(0,\dots,\underbrace{1}_{\text{i+1}},\dots,0)$ and $\epsilon^i_2=(0,\dots,\underbrace{-1}_{\text{i+1}},\dots,0)$. For an even number of dimensions, one may chose the basis in an equivalent way as an ensemble of two by two circular polarizations. In $D=6$ dimensions, for instance, this would give 
	\begin{align}
		\epsilon_1^1&=\frac{1}{\sqrt{2}}(0,-1,-i,0,0,0) \qquad \,\epsilon_2^1= \frac{1}{\sqrt{2}}(0,1,-i,0,0,0) \nonumber \\
		\epsilon_1^2&=\frac{1}{\sqrt{2}}(0,1,-i,0,0,0) \qquad \quad \epsilon_2^2=\frac{1}{\sqrt{2}}(0,-1,-i,0,0,0) \nonumber \\
		\epsilon_1^3&=\frac{1}{\sqrt{2}}(0,0,0,-1,-i,0) \qquad \,\epsilon_2^3= \frac{1}{\sqrt{2}}(0,0,0,1,-i,0) \nonumber \\
		\epsilon_1^4&=\frac{1}{\sqrt{2}}(0,0,0,1,-i,0) \qquad \quad \epsilon_2^4=\frac{1}{\sqrt{2}}(0,0,0,-1,-i,0).
	\end{align}
	Of course, the results are independent of the particular choice. 
	
	Whatever specific basis one choses, from \eqref{photon amplitude helicity} it follows that in the threshold limit, as already observed for the specific case $D=4$, only the diagonal terms $\mathcal M_{ii}$ are non zero, and they all give the same contribution
	\begin{equation}
		\mathcal M_{ii}\to 2(gq)^4 \left(1-\frac 12\frac{D-1}{D-2}\right).
	\end{equation} 
	It is then straightforward to extract the value of the amplitude in the threshold limit for $D$ generic dimensions as 
	\begin{equation}
		\label{photon pair production D generic}
		\left|\mathcal M\right|^2\to\frac{1}{(D-2)^2}(D-2)\left|\mathcal M_{ii}\right|^2=\frac{4}{D-2}(gq)^4\left(1-\frac 12\frac{D-1}{D-2}\right)^2=\left(\frac{D-3}{D-2}\right)^2\frac{(gq_n)^4}{D-2}.
	\end{equation}
	This result of course matches that shown in \eqref{photon photon pair production NR limit}, that was obtained by means of the usual trick $\sum_{\rm pol}\epsilon_\mu(p)\epsilon_\nu(p)=-g_{\mu\nu}$. Note also that when the dilaton is put to zero (i.e. when the second contribution in the parenthesis \eqref{photon pair production D generic} is put to wero) we re-obtain the result
	\begin{equation}
		\left|\mathcal M_{\gamma\gamma}\right|^2\to \frac{4}{D-2}(gq)^4.
	\end{equation}
	
	The same procedure can now be used to extract the different components of the purely gravitational amplitude of section \ref{subsection purely gravitational pair production}. The four diagrams contribute in the amount
	\begin{align}
		\mathcal M_{\rm t-pole}&=-\frac{4 \kappa ^2 (\epsilon_1\cdot p_3)^2 (\epsilon_2\cdot p_4)^2}{t-m_n^2} \nonumber \\
		\mathcal M_{\rm u-pole}&=-\frac{4 \kappa ^2 (\epsilon_2\cdot p_3)^2 (\epsilon_1\cdot p_4)^2}{u-m_n^2}  \\
		\mathcal M_{\rm seagull}&=2 \kappa ^2 \epsilon_1\cdot\epsilon_2 \left(\epsilon_1\cdot\epsilon_2 \left(p_3\cdot p_4+m_n^2\right)-2 \epsilon_2\cdot p_3\, \epsilon_1\cdot p_4-2\epsilon_1\cdot p_3 \, \epsilon_2\cdot p_4\right) \nonumber
	\end{align}
	and 
	\begin{align}
		\mathcal M_{\rm g-pole}=\frac{2\, \epsilon_1\cdot \epsilon_2}{D-2}\Bigg\{&2\, p_1\cdot p_2 \Big[(D-2) (\epsilon_{2,\lambda} \epsilon _{1,\tau}+\epsilon_{1,\lambda} \epsilon_{2,\tau})-\epsilon _1\cdot\epsilon_2\, \eta_{\lambda\tau} \Big]\nonumber\\
		&+p_1\cdot p_2 \Big[4\, \epsilon_1\cdot\epsilon_2\, \eta_{\lambda\tau}-2(D-2)(\epsilon_{2,\lambda}\epsilon_{1,\tau}+\epsilon_{1,\lambda}\epsilon_{2,\tau})\Big]\nonumber \\
		&+D \,\epsilon_1\cdot\epsilon_2\,  ( p_{1,\lambda} p_{1,\tau}+p_{2,\lambda} p_{2,\tau}+p_{1,\lambda}(p_1+p_2)_\tau+p_{2,\lambda}(p_1+p_2)_\tau) \nonumber\\
		&+2 D\, p_1\cdot p_2 \,\epsilon_{2,\lambda} \epsilon_{1,\tau}+2 (D-2)\, p_1\cdot p_2\, \epsilon_{1,\lambda} \epsilon_{2,\tau}-2 \,p_1\cdot p_2 \,\epsilon_1\cdot\epsilon_2 \eta_{\lambda\tau}\nonumber\\
		&+2\, \epsilon_1\cdot\epsilon_2\, p_{2,\lambda} p_{1,\tau}+2 \epsilon_1\cdot\epsilon_2\, p_{1,\lambda} p_{2,\tau}-2 \epsilon_1\cdot\epsilon_2\, (p_1+p_2)_\lambda p_{1,\tau}\nonumber\\
		&-2 \epsilon_1\cdot\epsilon_2\, (p_1+p_2)_\lambda p_{2,\tau}-2 \epsilon_1\cdot\epsilon_2\, p_{1,\lambda} (p_1+p_2)_\tau-2 \epsilon_1\cdot\epsilon_2\, p_{2,\lambda} (p_1+p_2)_\tau\nonumber\\
		&-4\, p_2\cdot (p_1+p_2)\, \epsilon_{2,\lambda} \epsilon_{1,\tau}\Bigg\}\Bigg(p^{3, \lambda} p^{4,\tau}+p^{4,\lambda} p^{3, \tau} -g^{\lambda\tau}\left(p_3\cdot p_4+m_n^2\right)\Bigg)	
	\end{align}
	to give \eqref{graviton graviton amplitude}, reported here for simplicity
	\begin{align*}
		i\mathcal M_{GG}=&\frac{\kappa ^2}{2} \left(-\frac{8 (p_3\cdot\epsilon_1)^2 (p_4\cdot\epsilon_2)^2}{t-m_n^2}-\frac{8 (p_3\cdot\epsilon_2)^2 (p_4\cdot\epsilon _1)^2}{u-m_n^2} \right.\nonumber\\
		&\left.\qquad-2\frac{(\epsilon_1\cdot\epsilon _2)^2 \left(m_n^4-tu-sm_n^2\right)}{s}-4 \epsilon_1\cdot\epsilon_2 \left(p_3\cdot\epsilon_2\, p_4\cdot\epsilon_1+p_3\cdot\epsilon_1\, p_4\cdot\epsilon_2\right)\right)
	\end{align*}
	As in the previous case, it is again easily verified that in the threshold limit only the diagonal $\mathcal M_{ii}$ terms are non-vanishing and that they all give the same result. In terms of the above amplitude, such non-vanishing contribution is given by the $(\epsilon_1\cdot\epsilon_2)^2$ term that results in
	\begin{equation}
		\mathcal{M}_{ii}\to \kappa^2m_n^2.	
	\end{equation}
	It is now straightforward to obtain, from these considerations, the result for the squared amplitude in $D$ generic dimensions: 
	\begin{equation}
		\left|\mathcal M\right|^2\to \frac{1}{(D-2)^2}(D-2)\left|\mathcal{M}_{ii}\right|^2=\frac{\kappa^4m_n^4}{D-2},
	\end{equation}
	which is the result quoted in the text \eqref{d dimensions gravity}.

	\providecommand{\href}[2]{#2}\begingroup\raggedright\endgroup
	
\end{document}